\documentclass[prd,aps,preprint,amsmath,nofootinbib,amssymb,eqsecnum,showkeys,tightenlines]{revtex4-1}
\usepackage{hyperref} % should be commented out if the tex file will be compiled with latex in arXiv!!! (pdflatex is fine)
\usepackage{slashed}
\usepackage{epsfig,latexsym,cancel,amssymb,amsmath,verbatim,mathrsfs}
\usepackage{color}
\usepackage{graphicx}
\usepackage{caption}
\usepackage{subcaption}
\usepackage[inline]{enumitem}
\usepackage[multidot]{grffile}  % allow figure file names to include dots
\usepackage{dcolumn}
\usepackage{bm}
\usepackage{amsmath}
\usepackage{amsfonts}
\usepackage{amssymb}
\usepackage{color}
\usepackage{latexsym}
\usepackage{slashed} % slash mark
\usepackage{pstricks}
\usepackage{indentfirst}
\usepackage{mathrsfs}
\usepackage{multirow}
\usepackage{epsfig,psfrag}
\usepackage{setspace} % spacing
\usepackage[utf8]{inputenc} % accept utf-8 input encoding
\usepackage[scientific-notation=true]{siunitx} % comprehensive units

\allowdisplaybreaks % allow eqnarray breaks
\newcommand{\beq}{\begin{eqnarray}}% can be used as {equation} or  {eqnarray}
\newcommand{\eeq}{\end{eqnarray}}

\graphicspath{{fig/}}

\allowdisplaybreaks

\begin{document}

	%%%%%%%%%%%%%%TITLE AFFILIATIONS ETC%%%%%%%%%%%%%%%%%%%%%%%%%%%%%%%%%%%%%%%%%%%%%%%%%%%%%%%%%%%%
	
	\title{\texorpdfstring{\Large Vector dark matter production from catalyzed annihilation}{}}
	\author{Chengfeng {\sc Cai}}
	\email{caichf3@mail.sysu.edu.cn}
	\affiliation{School of Physics, Sun Yat-Sen University, Guangzhou 510275, China}
	\author{Hong-Hao {\sc Zhang}}
	\email[Corresponding author. ]{zhh98@mail.sysu.edu.cn}
	\affiliation{School of Physics, Sun Yat-Sen University, Guangzhou 510275, China}
	%%%%%%%%%%%%%%%%%%%%%%%%%%%%%%%%%%%%%%%%%%%%%%%%%%%%%%%%%%%%%%%%%%%%%%%%%%
	
	%%%%%%%%%%%%%%%%%%%%%%%%%%%%%%%%%%%%%%%%%%%%%%%%%%%%%%
	%%%%%%%%%%%%%%%%%%%%%%%%%%%%%%%%%%%%%%%%%%%%%%%%%%%%%%
	\begin{abstract}
		We provide a simple model of vector dark matter (DM) which can realize the recently proposed freeze-out mechanism with catalyzed annihilation. In our setup, a vector DM field $X_\mu$ and a catalyst field $C_\mu$ is unified by an SU(2)$_D$ gauge symmetry. These gauge fields acquire their masses via spontaneously symmetry breaking triggered by a doublet and a real triplet scalar fields. The catalyst particle is automatically lighter than the DM since it only acquires mass from the vacuum expectation value of the doublet scalar. We also introduce a dimension-5 operator to generate a kinetic mixing term between $C_\mu$ and the U(1)$_Y$ gauge field $B_\mu$. This mixing term is naturally small due to a suppression with a high UV completion scale, and thus it allows the catalyst to decay after the DM freeze-out. We derive the annihilation cross sections of processes $X^\ast+X\to 2C$ and $3C\to X^\ast+X$ and solve the Boltzmann equations for both the DM and the catalyst. We develop the analytical approximate solutions of the equations and find them matching the numerical solutions well. Constraints from relic abundance and indirect detection of DM are considered. We find that the DM with a mass $m_X\gtrsim4.5$~TeV survives in the case of a long-living catalyst. On the other hand, if the catalyst decays during the catalyzed annihilation era, then the bound can be released. We also discuss two paradigms which can maintain the kinetic equilibrium of DM until the DM freeze-out. In both cases, the freeze-out temperature of DM is an order of magnitude higher than the original model.
		\\[.3cm]
	\end{abstract}
	\maketitle
	\newpage

	\section{Introduction}
	Dark Matter (DM) constitutes about 27\% of energy density in the Universe, but its particle properties and production mechanism remain still unknown to us. Observations from cosmology and astrophysics indicate that the DM is mostly likely to be cold when it decouples from the thermal bath. One of the most popular types of cold DM is Weakly Interacting Massive Particles (WIMPs), which are thermally produced in the early Universe and finally frozen out at some temperature $T_f\sim m_{DM}/25$. In this kind of models, DM candidates usually have masses ranging from $1$ GeV to $10$~TeV and the magnitude of their couplings with SM particles are similar to the weak interaction. Based on these implications, people have designed many experiments to detect  WIMPs directly~\cite{Akerib:2016vxi,Aprile:2018dbl,PandaX-4T:2021bab} and indirectly~\cite{Strigari:2013iaa,Adriani:2013uda,Ackermann:2015zua,Fermi-LAT:2016uux,Profumo:2017obk,Hoof:2018hyn}.
	
	Recently, a new DM freeze-out paradigm is proposed by Xing and Zhu in Ref.\cite{Xing:2021pkb}. In their setup, the dark sector is nearly secluded, and the depletion of a DM particle $\chi$ is assisted with a catalyst particle $A'$, which is slightly lighter than $\chi$. The dominant processes are $2\chi\to2A'$ and $3A'\to2\chi$, in which the yield of $A'$ ($Y_{A'}$) keeps nearly constant until $A'$ decays. Note that the model is similar to the secluded DM~\cite{Pospelov:2007mp,ArkaniHamed:2008qn}, but the lifetime of the catalyst particle is much longer. They are required to be long-living enough to support the whole catalyzed annihilation processes until the DM  freeze-out. In this way, the yield of DM decreases in a manner of $Y_\chi\propto x^{-3/2}$ during the catalyzed annihilation era. Comparing with the situation of Strongly Interacting Massive Particles (SIMPs) models and their variations~\cite{Hochberg:2014kqa,Hochberg:2014dra,Bernal:2015xba,Bernal:2017mqb,Smirnov:2020zwf}, the depletion efficiency of DMs in the catalyzed annihilation scenario is much slower, and thus the freeze-out temperature is lower. In Ref.\cite{Xing:2021pkb}, an $\textrm{U}(1)'$ gauge symmetric model with fermionic DM is presented to illustrate how the catalyzed freeze-out mechanism does work. A tiny kinetic mixing between the dark photon and the U(1)$_Y$ gauge field is introduced to enable the catalyst decay.
	
	In this work, we propose a vector DM model in which the DM candidates freeze-out through the catalyzed annihilation. Vector dark matter models has been discussed in many previous studies, such as a U(1) gauge symmetry extension~\cite{Farzan:2012hh, Abe:2012efa, Arcadi:2020jqf, Baek:2012se,Chen:2014cbt,DiFranzo:2015nli, LEBEDEV2012570, duch_stable_2015, Hadjimichef:2016zsk,Zaazoua:2021xls,Adshead:2021kvl}, a non-abelian gauge symmetry extension~\cite{Hambye:2008bq, Hambye:2009fg,SU2DMDR,Davoudiasl:2013jma,Gross:2015cwa, Chen:2015nea,Karam:2015jta,Karam:2016rsz,Ko:2016fcd,Choi:2017zww,Abe:2020mph,Hisano:2020qkq,Hu:2021pln,Baouche:2021wwa} and a model with non-gauge field vector bosons~\cite{Belyaev:2018xpf}. We consider an SU(2)$_D$ gauge symmetry which is spontaneously broken by a doublet scalar $\Phi_D^i$ and a real triplet scalar $\Delta_D^a$. A complex vector field $X_\mu\equiv (V^1_\mu-i V^2_\mu)/\sqrt{2}$, which is formed by two components of the  SU(2)$_D$ gauge fields, is regarded as a DM candidate. The remaining gauge field $C_\mu\equiv V^3_\mu$ plays the role of a catalyst. It means that the DM and the catalyst are unified in our model. In order to allow the catalyst to decay, we introduce a dimension-5 effective operator $B^{\mu\nu}\Delta_D^aV^a_{\mu\nu}$ which generates a kinetic mixing term between the catalyst field $C_\mu$ and the U(1)$_Y$ gauge field $B_\mu$~\cite{Holdom:1985ag}. This kinetic mixing term can be naturally small since the operator can be suppressed by a large UV completion scale. A condition of catalyzed annihilation is that the catalyst should be lighter than the DM. It is automatically satisfied in our setup since $C_\mu$ only acquires mass from the vacuum expectation value (VEV) of the doublet scalar while $X_\mu$ acquires mass from both VEVs of the doublet and the triplet. The processes of DMs annihilating into catalysts can lead to significant signals in DM indirect detection experiments, such as the Fermi-LAT~\cite{Hoof:2018hyn} and the CTA~\cite{Doro:2012xx}. We will discuss their constraints and sensitivities in our model.
	
    In the framework of catalyzed freeze-out, a tough problem is raised that the interactions between the dark and the SM sectors are too weak to keep the DM in kinetic equilibrium (KE) with the thermal bath during the catalyzed annihilation era. We propose two template models to alleviate this problem. One is to maintain KE via the DM-fermions scattering mediated by Higgs bosons. It requires a larger Higg-portal coupling comparing to the original model. The other one is to maintain the KE assisted by a thermal axion-like particle (ALP). In both cases, at least one more parameter is needed for determining the freeze-out temperature of the catalyst.
	
	This paper is organized as follows. In section \ref{sect2}, we introduce the SU(2)$_D$ gauge models. In section \ref{sect3}, we discuss the solutions of the Boltzmann equations and some relevant constraints from experiments. In section \ref{sect4}, we discuss two strategies to solve the kinetic equilibrium problem. Finally, we conclude all our findings in the section \ref{sect5}.

	\section{The model}\label{sect2}
	\subsection{SU(2)$_D$ gauge-Higgs model}
	In this section, let us present the model. We extend the SM with an SU(2)$_D$ gauge symmetry which is spontaneously and completely broken by a scalar doublet and a triplet. All the three components of the gauge fields will be massive and two of them are degenerate. The degenerate components can combine to form a complex vector field $X_\mu$ (similar to the W boson in the SM), which is charged under a global U(1)$_D$ symmetry, while SM particles are neutral. If $X_\mu$ is the lightest particle with U(1)$_D$ charge, then it can be a stable DM candidate since it does not completely decay into the SM particles. The remaining component of the SU(2)$_D$ gauge fields is a real vector field $C_\mu$, which is lighter than $X_\mu$ and thus it can play the role of a catalyst.
	
	The Lagrangian of the pure gauge part is
	\beq
	\mathcal{L}_{gauge}=-\frac{1}{4}V^a_{\mu\nu}V^{a\mu\nu},\label{eq-1}
	\eeq
	where $V^a_{\mu\nu}=\partial_\mu V^a_\nu-\partial_\nu V^a_\mu+g_D\epsilon^{abc}V^b_\mu V^c_\nu$ is the field strength tensor of the SU(2)$_D$ gauge fields $V^a_{\mu}$ ($a=1,2,3$) with gauge coupling $g_D$.
	Let us denote $X_\mu\equiv(V^1_\mu-iV^2_\mu)/\sqrt{2}$ and  $C_\mu\equiv V^3_\mu$, and rewrite the Lagrangian \eqref{eq-1} as
	\beq\label{Lgauge}
	\mathcal{L}_{gauge}&=&-\frac{1}{4}C_{\mu\nu} C^{\mu\nu}-\frac{1}{2}\hat{X}^\ast_{\mu\nu}\hat{X}^{\mu\nu}-g_D^2(C_\mu C^\mu X^\nu X^{\ast}_\nu-C_\mu C_\nu X^\mu X^{\ast\nu})\nonumber\\
	&&-\frac{ig_D}{2}\hat{X}^{\mu\nu}(C_\mu X^\ast_\nu-C_\nu X^\ast_\mu)+\frac{ig_D}{2}\hat{X}^{\ast\mu\nu}(C_\mu X_\nu-C_\nu X_\mu)\nonumber\\
	&&+ig_DC^{\mu\nu}X_\mu X^\ast_\nu-\frac{g_D^2}{2}[(X^\ast_\mu X^\mu)^2-(X_\mu X^\mu)(X^\ast_\nu X^{\ast\nu})],
	\eeq
	where $C_{\mu\nu}\equiv\partial_\mu C_\nu-\partial_\nu C_\mu$ and $\hat{X}_{\mu\nu}\equiv\partial_\mu X_\nu-\partial_\nu X_\mu$.
	To generate the masses of the vector fields, we introduce an SU(2)$_D$ doublet scalar $\Phi_D^i=(\phi_1,\phi_2)$ and a real triplet scalar $\Delta_D^a=(\Delta_D^1,\Delta_D^2,\Delta_D^3)$. The gauge fields couple to the Higgs fields through the covariant derivative terms:
	\beq\label{Hcov}
	\mathcal{L}_H= (D_\mu \Phi_D)^\dag D^\mu \Phi_D+ \mathrm{tr}[(D_\mu \Delta_D)^\dag D^\mu \Delta_D],
	\eeq
	where $\Delta_D\equiv\Delta_D^a\sigma^a/2$ with the Pauli matrices $\sigma^a$. The covariant derivatives of the scalar fields are given by
	\beq
	D_\mu \Phi_D&=&\left[\partial_\mu-ig_D\begin{pmatrix}\frac{C_\mu}{2}&\frac{X_\mu}{\sqrt{2}}\\ \frac{X^\ast_\mu}{\sqrt{2}}&-\frac{C_\mu}{2}\end{pmatrix}\right]\begin{pmatrix}\phi_1\\ \phi_2\end{pmatrix},\\
	D_\mu \Delta_D&=&\partial_\mu \begin{pmatrix}\frac{\Delta_D^3}{2}&\frac{\Delta}{\sqrt{2}}\\ \frac{\Delta^\ast}{\sqrt{2}}&-\frac{\Delta_D^3}{2}\end{pmatrix}-ig_D\left[\begin{pmatrix}\frac{C_\mu}{2}&\frac{X_\mu}{\sqrt{2}}\\ \frac{X^\ast_\mu}{\sqrt{2}}&-\frac{C_\mu}{2}\end{pmatrix},\begin{pmatrix}\frac{\Delta_D^3}{2}&\frac{\Delta}{\sqrt{2}}\\ \frac{\Delta^\ast}{\sqrt{2}}&-\frac{\Delta_D^3}{2}\end{pmatrix}\right]~,
	\eeq
	where we have defined a complex scalar field $\Delta\equiv(\Delta_D^1-i\Delta_D^2)/\sqrt{2}$ for convenience.
	To trigger the spontaneous breaking of SU(2)$_D$, we let the scalar fields acquire non-zero vacuum expectation values (VEVs), and parametrize them as
	\beq\label{param_H}
	\Phi_D=\begin{pmatrix}\phi_1\\ \frac{v_2+\varphi+ia}{\sqrt{2}}\end{pmatrix},\quad  \Delta_D=\begin{pmatrix}\frac{v_3+\rho}{2}&\frac{\Delta}{\sqrt{2}}\\ \frac{\Delta^\ast}{\sqrt{2}}&-\frac{v_3+\rho}{2}\end{pmatrix},
	\eeq
	where $v_2/\sqrt{2}$ and $v_3$ are the VEVs of $\phi_2$ and $\Delta_D^3$, respectively. Substituting eq.\eqref{param_H} into eq.\eqref{Hcov}, we obtain
	\beq\label{covder_explicit}
	\mathcal{L}_H&=&\frac{1}{2}(\partial_\mu\varphi)^2+\frac{1}{2}(\partial_\mu a)^2+(\partial_\mu\phi_1)^\ast\partial^\mu\phi_1+\frac{1}{2}(\partial_\mu\rho)^2+(\partial_\mu\Delta)^\ast\partial^\mu\Delta\nonumber\\
	&&+\frac{g_D}{2}C_\mu\left(\varphi\overleftrightarrow{\partial^\mu} a+\phi_1^\ast i\overleftrightarrow{\partial^\mu}\phi_1+2\Delta^\ast i\overleftrightarrow{\partial^\mu}\Delta\right)\nonumber\\
	&&+\frac{g_D}{2}X_\mu \left(\phi_1^\ast i\partial^\mu\varphi-\phi_1^\ast\overleftrightarrow{\partial^\mu} a-2\Delta^\ast\overleftrightarrow{\partial^\mu}\rho\right)+h.c.\nonumber\\
	&&+g_D^2C_\mu C^\mu\left[\frac{v_2^2}{8}+\frac{v_2}{4}\varphi+\frac{1}{8}\varphi^2+\frac{1}{4}|\phi_1|^2+\frac{1}{8}a^2+|\Delta|^2\right]\nonumber\\
	&&+g_D^2X^\ast_\mu X^\mu\left[\frac{v_2^2}{4}+v_3^2+\frac{v_2}{2}\varphi+2v_3\rho+\frac{1}{4}\varphi^2+\frac{1}{2}|\phi_1|^2+\frac{1}{4}a^2+\rho^2+|\Delta|^2\right]\nonumber\\
	&&-g_D^2(v_3+\rho)X^\ast_\mu C^\mu\Delta+h.c..
	\eeq
	The masses of gauge fields are found to be
	\beq
	m_C=\frac{g_D}{2}v_2,\quad m_X=\frac{g_D}{2}\sqrt{v_2^2+4v_3^2}\equiv\frac{g_D}{2}v_1~,
	\eeq
	where we have defined $v_1\equiv\sqrt{v_2^2+4v_3^2}$. It is obvious that $X_\mu$ is heavier than $C_\mu$ due to the contribution from $v_3$. If $v_3\lesssim0.56v_2$ ($1.5m_C\gtrsim m_X$),  then the  annihilation process $3C\to X+X^\ast$ can happen in the non-relativistic limit.
	
	To justify the vacuum configuration, we need to figure out the minimum of the following potential terms of the scalar fields:
	\beq\label{potential}
	V&=&-\mu^2 |H|^2+\frac{\lambda}{2}|H|^4-\mu_2^2|\Phi_D|^2+\frac{\lambda_2}{2}|\Phi_D|^4-\mu_3^2 \mathrm{tr}[\Delta_D^\dag \Delta_D]+\frac{\lambda_3}{2}( \mathrm{tr}[\Delta_D^\dag \Delta_D])^2\nonumber\\
	&&+\lambda_{23}|\Phi_D|^2\mathrm{tr}[\Delta_D^\dag \Delta_D]+\kappa_{23}\Phi_D^\dag\Delta_D\Phi_D+\lambda_{02}|H|^2|\Phi_D|^2+\lambda_{03}|H|^2\mathrm{tr}[\Delta_D^\dag \Delta_D]
	\eeq
	where $H$ is the SM Higgs field parametrized as $H=(G^+,(v+h+i\chi)/\sqrt{2})^T$.
	The extremum conditions of the potential are
	\beq
	&&\left[-\mu^2+\frac{\lambda}{2}v^2+\frac{1}{2}(\lambda_{02}v_2^2+\lambda_{03}v_3^2)\right]v=0,\\
	&&\left[-\mu_2^2+\frac{\lambda_2}{2}v_2^2+\frac{\lambda_{23}}{2}v_3^2+\frac{1}{2}\lambda_{02}v^2-\frac{\kappa_{23}}{2}v_3\right]v_2=0,\\
	&&\left[-\mu_3^2+\frac{\lambda_3}{2}v_3^2+\frac{\lambda_{23}}{2}v_2^2+\frac{1}{2}\lambda_{03}v^2-\frac{\kappa_{23}}{2}\frac{v_2^2}{v_3}\right]v_3=0~.
	\eeq
	The mass matrix of the neutral CP-even fields in $(\varphi,\rho,h)$ basis is given by
	\beq\label{masseven}
	M_{even}^2=\begin{pmatrix}\lambda_2v_2^2&(\lambda_{23}-\xi_{23})v_2v_3&\lambda_{02}vv_2\\ (\lambda_{23}-\xi_{23})v_2v_3&\lambda_3v_3^2+\frac{1}{2}\xi_{23}v_2^2&\lambda_{03}vv_3\\\lambda_{02}vv_2&\lambda_{03}vv_3&\lambda v^2\end{pmatrix},
	\eeq
	where $\xi_{23}\equiv\kappa_{23}/2v_3$. It can be diagonalized by a orthogonal $3\times 3$ matrix $O$ as follows,
	\beq
	M_{diag}^2=OM_{even}^2 O^T=\mathrm{diag}\{m_3^2,m_2^2,m_1^2\}.
	\eeq
	We assume $\lambda_{02}$ and $\lambda_{03}$ to be much smaller than $\lambda_2$ and $\xi_{23}$ for obtaining a SM-like Higgs boson. The smallness of $\lambda_{02}$ and $\lambda_{03}$ also suppresses the annihilation cross sections of processes, $X^\ast+X\to \bar{t}+t, W^++W^-, Z+Z$, through Higgs-portal. With this assumption, the orthogonal matrix $O$ can now be approximated by
	\beq
	O\approx\begin{pmatrix}1&0&-\alpha_{13}\\0&1&-\alpha_{23}\\ \alpha_{13}&\alpha_{23}&1\end{pmatrix}\begin{pmatrix}c_\alpha&-s_\alpha&0\\s_\alpha&c_\alpha&0\\0&0&1\end{pmatrix}~,
	\eeq
	where $s_\alpha\equiv \sin\alpha$ and $c_\alpha\equiv\cos\alpha$ and
	\beq
	&&\tan(2\alpha)=\frac{2(\lambda_{23}-\xi_{23})v_2v_3}{\lambda_2v_2^2-\lambda_3v_3^2-\frac{\xi_{23}}{2}v_2^2}~,\\
	&&\alpha_{13}\approx-\frac{(\lambda_{02}v_2c_\alpha-\lambda_{03}v_3s_\alpha )v}{\lambda_2v_2^2c_\alpha^2+\left(\lambda_3v_3^2+\frac{\xi_{23}}{2}v_2^2\right)s_\alpha^2-(\lambda_{23}-\xi_{23})v_2v_3s_{2\alpha}-\lambda v^2}~,\label{alpha13}\\
	&&\alpha_{23}\approx-\frac{(\lambda_{02}v_2s_\alpha+\lambda_{03}v_3c_\alpha )v}{\lambda_2v_2^2s_\alpha^2+\left(\lambda_3v_3^2+\frac{\xi_{23}}{2}v_2^2\right)c_\alpha^2+(\lambda_{23}-\xi_{23})v_2v_3s_{2\alpha}-\lambda v^2}~\label{alpha23}.
	\eeq
	The mass eigenstates and corresponding eigenvalues are given by
	\beq
	\begin{pmatrix}h_3\\h_2\\h_1\end{pmatrix}&=&O\begin{pmatrix}\varphi\\ \rho\\ h\end{pmatrix}\\
	m_3^2&\approx&\lambda_2v_2^2c_\alpha^2+\left(\lambda_3v_3^2+\frac{\xi_{23}}{2}v_2^2\right)s_\alpha^2-(\lambda_{23}-\xi_{23})v_2v_3s_{2\alpha}\\
	m_2^2&\approx&\lambda_2v_2^2s_\alpha^2+\left(\lambda_3v_3^2+\frac{\xi_{23}}{2}v_2^2\right)c_\alpha^2+(\lambda_{23}-\xi_{23})v_2v_3s_{2\alpha}\\
	m_1^2&\approx&\lambda v^2
	\eeq
	We will assume that $m_1\approx126$~GeV is the mass of SM-like Higgs boson in later discussion.
	The CP-odd scalar $a$ is a Goldstone boson eaten by the gauge field $C_\mu$. The mass matrix of the complex scalar $(\phi_1,\Delta)$ is
	\beq
	M_{c}^2=\frac{1}{2}\xi_{23}\begin{pmatrix}4v_3^2&2v_2v_3\\ 2v_2v_3&v_2^2\end{pmatrix}~,
	\eeq
	which can be diagonalized by a rotation
	\beq
	R_\theta M_{c}^2 R_\theta^T=\begin{pmatrix}0&0\\0&m_s^2\end{pmatrix},\quad R_\theta=\begin{pmatrix}c_\theta&-s_\theta\\ s_\theta&c_\theta\end{pmatrix},
	\eeq
	where
	\beq
	s_\theta\equiv\sin\theta=\frac{2v_3}{v_1},\quad c_\theta\equiv\cos\theta=\frac{v_2}{v_1},\quad m_s^2=\xi_{23}v_1^2~.
	\eeq
	\subsection{Dimension-5 effective operator}
	We can check that \eqref{Lgauge}, \eqref{covder_explicit}, and \eqref{potential} are invariant under a global U(1)$_D$ transformation:
	\beq
	\Delta\to e^{i\gamma}\Delta,\quad \phi_1\to e^{i\gamma}\phi_1,\quad X_\mu\to e^{i\gamma}X_\mu.
	\eeq
	Therefore, $X_\mu$ cannot decay if it is lighter than $\Delta$ and $\phi_1$.
	In addition, there is a discrete symmetry $G_D$ in the SU(2)$_D$ gauge and Higgs sector. We can check that \eqref{Lgauge}, \eqref{Hcov} and \eqref{potential} are invariant under the following $G_D$ transformations
	\beq
	C_\mu\to-C_\mu,\quad X_\mu\to X^\ast_\mu,\quad\phi_1\to-\phi_1^\ast,\quad \phi_2\to\phi_2^
	\ast\quad \Delta\to-\Delta^\ast,\quad \Delta_D^3\to\Delta_D^3.
	\eeq
	This symmetry is preserved even after the $\phi_2$ and $\Delta_D^3$ acquiring VEVs.
	If we assume that the gauge fields $C_\mu$ and $X_\mu$ are much lighter than the Higgs fields $\Delta_D$ and $\Phi_D$, then the lightest particle in this sector is $C_\mu$ and it can not decay either due to the $G_D$ symmetry. According to the requirement of the catalyzed freeze-out mechanism, the catalyst $C_\mu$ should be long-living but unstable, so we need to add something new to slightly violate $G_D$. As an effective theory in low energy, we can introduce a dimension-5 operator:
	\beq\label{effop}
	\mathcal{L}_{5}=-\frac{c}{\Lambda}B^{\mu\nu}\Delta_D^a V^a_{\mu\nu},
	\eeq
    where $c$ is a Wilson coefficient, and $\Lambda $ is some UV complete scale. We can check that $\Delta_D^a V^a_{\mu\nu}\to -\Delta_D^a V^a_{\mu\nu}$ under the $G_D$ transformation, and thus $\mathcal{L}_{5}$ violates the symmetry. Substituting \eqref{param_H} into the operator \eqref{effop}, we find it includes the following terms,
    \beq
    \mathcal{L}_5\supset-\frac{c}{\Lambda}B^{\mu\nu}\Delta_D^3 V^3_{\mu\nu}=-\frac{c(v_3+\rho)}{\Lambda}B^{\mu\nu}C_{\mu\nu}+\frac{ig_Dc(v_3+\rho)}{\Lambda}B^{\mu\nu}(X_\mu X^\ast_\nu-X^\ast_\mu X_\nu)~.
    \eeq
    The first term is an effective kinetic mixing between $B_\mu$ and $C_\mu$ fields, while the second term includes an electromagnetic interaction of the magnetic moment of $X_\mu$\footnote{In Ref.\cite{Hisano:2020qkq}, the electric and magnetic multipole moments of vector DM are studied in details.}. Due to the kinetic mixing, the $C_\mu$ can finally decay into SM particles.

    Note that the kinetic terms of $B_\mu$ and $C_\mu$ are not in the canonical form, so we should figure out a new basis $(\hat{B}_\mu,\hat{C}_\mu)$ such that all fields are canonically normalized. It can be done by the following transformation of basis~\cite{Babu:1997st,Frandsen:2011cg,Lao:2020inc}:
    \beq
    \begin{pmatrix}B_\mu\\ C_\mu\end{pmatrix}= \begin{pmatrix}1&-t_\epsilon\\ 0&\frac{1}{c_\epsilon}\end{pmatrix}   \begin{pmatrix}\hat{B}_\mu\\ \hat{C}_\mu\end{pmatrix}~,
    \eeq
    where $s_\epsilon\equiv2cv_3/\Lambda$. The interaction part of the effective operator in terms of $(\hat{B}_\mu,\hat{C}_\mu)$ is given by
    \beq
    \mathcal{L}_{5}&\supset&-\frac{t_\epsilon}{2v_3}\rho\hat{B}^{\mu\nu}\hat{C}_{\mu\nu}+\frac{ig_Ds_\epsilon}{2v_3}\rho\hat{B}^{\mu\nu}(X_\mu X^\ast_\nu-X^\ast_\mu X_\nu)\nonumber\\
    &&-\frac{s_\epsilon}{2v_3}\hat{B}^{\mu\nu}\left[\Delta\left(  \hat{X}^\ast_{\mu\nu} +\frac{ig_D}{c_\epsilon}(\hat{C}_{\mu}X^\ast_\nu-\hat{C}_\nu X^\ast_\mu)\right)+h.c.\right]+...,
    \eeq
    where we have neglected $\mathcal{O}(s_\epsilon^2)$ and other higher order terms. In the new basis, the covariant derivatives of the scalar fields are given by
    \beq
    D_\mu H&=&\left[\partial_\mu-ig W^a_\mu\frac{\tau^a}{2}-i\frac{g'}{2}\hat{B}_\mu+\frac{ig't_\epsilon}{2}\hat{C}_\mu\right]H~,\\
    D_\mu\Phi_D&=&\partial_\mu\Phi_D-ig_D\begin{pmatrix}\frac{\hat{C}_\mu}{2c_\epsilon}&\frac{X_\mu}{\sqrt{2}}\\\frac{X^\ast_\mu}{\sqrt{2}}&-\frac{\hat{C}_\mu}{2c_\epsilon}\end{pmatrix}\Phi_D~,\\
    D_\mu\Delta_D&=&\partial_\mu\Delta_D-ig_D\left[\begin{pmatrix}\frac{\hat{C}_\mu}{2c_\epsilon}&\frac{X_\mu}{\sqrt{2}}\\\frac{X^\ast_\mu}{\sqrt{2}}&-\frac{\hat{C}_\mu}{2c_\epsilon}\end{pmatrix}\Delta_D-\Delta_D\begin{pmatrix}\frac{\hat{C}_\mu}{2c_\epsilon}&\frac{X_\mu}{\sqrt{2}}\\\frac{X^\ast_\mu}{\sqrt{2}}&-\frac{\hat{C}_\mu}{2c_\epsilon}\end{pmatrix}\right]~.
    \eeq
    The masses of $W^\pm_\mu$, $X_\mu$ and the neutral gauge fields $(W^3_\mu,\hat{B}_\mu,\hat{C}_\mu)$ can be read off as follows,
    \beq
    &&m_W^2=\frac{g^2}{4}v^2~,\quad m_X^2=g_D^2\left(\frac{v_2^2}{4}+v_3^2\right)~,\\ &&M_{g}^2=\frac{1}{4}\begin{pmatrix}g^2v^2&-gg'v^2&gg't_\epsilon v^2\\-gg'v^2&g'^2v^2&-g'^2t_\epsilon v^2\\gg't_\epsilon v^2&-g'^2t_\epsilon v^2&g'^2t_\epsilon^2v^2+g_D^2v_2^2\end{pmatrix}~.
    \eeq
    $M_g^2$ can be diagonalized by an orthogonal transformation $m_g^2=O_gM_g^2O_g^T=\mathrm{diag}\{0,m_Z^2,m_{Z'}^2\}$, where
    \beq
    &&O_g=\begin{pmatrix}1&0&0\\ 0&c_\zeta&-s_\zeta\\0&s_\zeta&c_\zeta\end{pmatrix} \begin{pmatrix}\hat{s}_W&\hat{c}_W&0\\ \hat{c}_W&-\hat{s}_W&0\\0&0&1\end{pmatrix}~,\\
    &&\tan(2\zeta)=\frac{2s_\epsilon c_\epsilon \hat{s}_W(g^2+g'^2)v^2}{(g^2+g'^2)v^2c_\epsilon^2(1-\hat{s}_W^2t_\epsilon^2)-g_D^2v_2^2}~,\\
    &&m_Z^2=\frac{(g^2+g'^2)}{4}(1+\hat{s}_Wt_\epsilon t_\zeta),\quad m_{Z'}^2=\frac{g_D^2v_2^2}{4c_\epsilon^2(1+\hat{s}_Wt_\epsilon t_\zeta)}~,
    \eeq
    and $\hat{s}_W\equiv\sin\hat{\theta}_W=g'/\sqrt{g^2+g'^2}$ is the sine of the Weinberg angle. When $s_\epsilon\ll1$ and $\sqrt{g^2+g'^2}v\ll g_Dv_2$, $t_\zeta\equiv \tan\xi$ can be approximated by
    \beq
    t_\zeta\approx\frac{\hat{s}_Wt_\epsilon}{1-r}~,
    \eeq
    where $r\equiv m_{Z'}^2/m_Z^2$. The mass eigenstate $Z'_\mu$ is the true catalyst particle and it is very closed to the gauge eigenstate $C_\mu$ in the $s_\epsilon\ll1$ limit.
    For discussing the phenomenologies later, we show the SM neutral current interactions in terms of gauge fields mass eigenstates as follows,
    \beq
    \mathcal{L}_{NC}&=&eJ_{EM}^\mu A_\mu+\left[\frac{g}{2\hat{c}_W}(\hat{s}_W s_\zeta t_\epsilon+c_\zeta)J_Z^\mu-e\hat{c}_W s_\zeta t_\epsilon J_{EM}^\mu\right]Z_\mu\nonumber\\
    &&+\left[\frac{g}{2\hat{c}_W}(\hat{s}_W c_\zeta t_\epsilon-s_\zeta)J_Z^\mu-e\hat{c}_W c_\zeta t_\epsilon J_{EM}^\mu\right]Z'_\mu,
    \eeq
    where $J_{EM}^\mu$ and $J_Z^\mu$ correspond to the neutral currents of SM fermions\footnote{More details can be found in Ref.\cite{Lao:2020inc}.}.

    Finally, we want to point out that a possible UV completion of the operator \eqref{effop} is to introduce a super heavy vector-like fermion $\Psi=(\Psi_1,\Psi_2)^T$ which is a doublet of SU(2)$_D$ with hypercharge $Y=-1$. The Lagrangian of $\Psi$ is given by
    \beq
    \mathcal{L}_\Psi=\bar{\Psi}(i\slashed{D}-m_\Psi)\Psi-y_3\bar{\Psi}\Delta_D\Psi-y_2^I\bar{\Psi}\Phi_D e_R^I+h.c.~,
    \eeq
    where $e_R^I$ is the $I$-th generation of right-handed charged lepton. When $y_3=y_2^I=0$, the $G_D$ symmetry is respected if $\Psi$ transforms in the following way
    \beq
    \Psi_1\leftrightarrow \Psi_2~.
    \eeq
    Once $y_3$ and $y_2^I$ are turned on, the $G_D$ symmetry is broken and then the operator \eqref{effop} can be induced by loops of $\Psi_{1,2}$. Using the formula given in Ref.\cite{Hisano:2020qkq}, the mixing parameter is
    \beq
    s_\epsilon\sim \frac{g_Dg'}{6\pi^2}\left(\frac{y_3v_3}{m_\Psi}\right)~.
    \eeq
    For the purpose of obtaining a value $t_\epsilon\sim 10^{-11}$, we need to set $m_\Psi\sim10^{12}$~GeV when $v_3\sim1$~TeV.
    
    \section{Catalyzed freeze-out of $X_\mu$}\label{sect3}
    \subsection{Annihilation cross sections and decay width}
    The dominant annihilation processes of DM pairs to SM particles are $X^\ast+X\to W^++W^-,~Z+Z,~h+h,~\bar{f}+f$ through s-channel mediated by Higgs bosons and gauge bosons. Since the annihilation cross sections of gauge boson portal processes are suppressed by $s_\epsilon^2\sim v_3^2/\Lambda^2$ which is assumed to be extremely small, we only need to compute the contributions from Higgs-portal processes. The corresponding annihilation cross sections are given by
    \beq
    \langle\sigma v\rangle_{X^\ast X\to 2\phi_{SM}}&\approx&\langle\sigma v\rangle_{X^\ast X\to \bar{t}t}+\langle\sigma v\rangle_{X^\ast X\to W^+W^-}+\langle\sigma v\rangle_{X^\ast X\to ZZ}+\langle\sigma v\rangle_{X^\ast X\to h_1h_1}\\
    \langle\sigma v\rangle_{X^\ast X\to \bar{t}t}&\approx&\frac{g_D^4v_2^2}{256\pi m_X^2v^2}\left(\frac{m_t^2}{m_X^2}\right)\left[(c_\alpha\alpha_{13}+s_\alpha\alpha_{23})+\frac{4v_3}{v_2}(c_\alpha\alpha_{23}-s_\alpha\alpha_{13})\right]^2\nonumber\\
    \langle\sigma v\rangle_{X^\ast X\to VV}&\approx&\frac{g_D^4v_2^2}{256\pi m_X^2v^2}\left[(c_\alpha\alpha_{13}+s_\alpha\alpha_{23})+\frac{4v_3}{v_2}(c_\alpha\alpha_{23}-s_\alpha\alpha_{13})\right]^2\nonumber\\
    \langle\sigma v\rangle_{X^\ast X\to h_1h_1}&\approx&\left(\frac{1}{3}\right)\frac{g_D^4v_2^2}{256\pi m_X^2v^2}\left[\alpha_{13}c_{2\alpha}+\left(\frac{m_2^2}{m_3^2}\right)\alpha_{23}s_{2\alpha}-\frac{4v_3}{v_2}\left(\alpha_{13}s_{2\alpha}-\left(\frac{m_2^2}{m_3^2}\right)\alpha_{23}c_{2\alpha}\right)\right]^2\nonumber\\
    \langle\sigma v\rangle_{Z' Z'\to 2\phi_{SM}}&\approx&\langle\sigma v\rangle_{Z'Z'\to \bar{t}t}+\langle\sigma v\rangle_{Z'Z'\to W^+W^-}+\langle\sigma v\rangle_{Z'Z'\to ZZ}+\langle\sigma v\rangle_{Z'Z'\to h_1h_1},\label{anntoSM}\\
    \langle\sigma v\rangle_{Z'Z'\to \bar{t}t}&\approx&\frac{g_D^4v_2^2}{512\pi m_{Z'}^2v^2}\left(\frac{m_t^2}{m_{Z'}^2}\right)(c_\alpha\alpha_{13}+s_\alpha\alpha_{23})^2\nonumber\\
     \langle\sigma v\rangle_{Z'Z'\to VV}&\approx&\frac{g_D^4v_2^2}{512\pi m_{Z'}^2v^2}(c_\alpha\alpha_{13}+s_\alpha\alpha_{23})^2\nonumber\\
     \langle\sigma v\rangle_{Z'Z'\to h_1h_1}&\approx&\left(\frac{1}{3}\right)\frac{g_D^4v_2^2}{256\pi m_{Z'}^2v^2}\left[\alpha_{13}c_{2\alpha}+\left(\frac{m_2^2}{m_3^2}\right)\alpha_{23}s_{2\alpha}\right]^2,\nonumber
    \eeq
    where $2\phi_{SM}$ represents all dominant SM final states including $\bar{t}t$, $W^+W^-$, $ZZ$, $h_1h_1$, while $VV$ represents two vector bosons $Z+Z$, $W^++W^-$. For simplicity, we have assumed that the model parameters satisfy $m_t^2,m_1^2,m_{W}^2,m_{Z}^2\ll m_X^2$ and $m_X^2,m_{Z'}^2\ll m_{2}^2,m_{3}^2$. All these annihilation cross sections are suppressed by the couplings, $\lambda_{02}$ and $\lambda_{03}$, and hence we can further assume $\lambda_{02}$ and $\lambda_{03}$ are very small such that $Z'+Z' (X^\ast+X)\to h_i\to 2\phi_{SM}$ processes fall behind the Hubble expansion early ($x\sim 10$). Note that very small $\lambda_{02}$ and $\lambda_{03}$ also suppress the Higgs-portal DM-nuclei scattering cross section, and thus the model can easily circumvent the stringent direct detection bound. However, a very weak coupling is hard to keep DM in kinetic equilibrium with the thermal bath until freeze-out. At the moment, we just assume that the kinetic equilibrium is maintained by some unknown mechanisms. In the next section, we will introduce two strategies for solving this problem, but we need to pay the price that one more parameter is needed for determining the relic abundance of DM.

\begin{figure}[tb]
    	\centering
    	\includegraphics[width=0.32\textwidth]{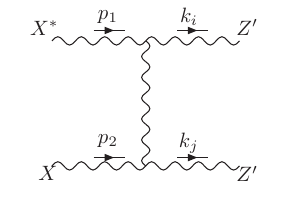}
    	\includegraphics[width=0.32\textwidth]{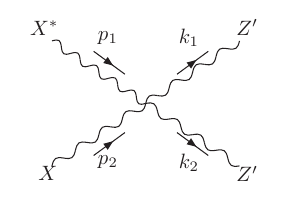}
    	\caption{Feynman diagrams of $X^\ast+X\to Z'+Z'$ processes. There are $2$ independent diagrams with $i,j=1,2$ and $i\neq j$ for the first plot.}
    	\label{diagram2to2}
    \end{figure}

    The annihilation cross section of $X^\ast+X\to Z'+Z'$ process is neither suppressed by the kinetic mixing parameter $s_\epsilon$ nor the Higgs mixing couplings $\lambda_{02},\lambda_{03}$. In the situation of $m_X^2\ll m_{2}^2,m_{3}^2$, the dominant diagrams of the process are shown in Fig.\ref{diagram2to2}, and the annihilation cross section to the leading order is
    \beq
    \langle\sigma_2 v\rangle\approx \frac{g_D^4(1-r_{XZ'}^{-1})^{1/2}(152r_{XZ'}^4-136r_{XZ'}^3+128r_{XZ'}^2-18r_{XZ'}+3)}{144\pi m_{Z'}^2r_{XZ'}^3(2r_{XZ'}-1)^2}~,
    \eeq
    where $r_{XZ'}=m_X^2/m_{Z'}^2\approx c_\theta^{-2}$.

\begin{figure}[tb]
    	\centering
    	\includegraphics[width=0.32\textwidth]{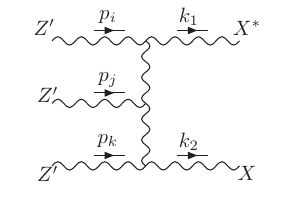}
    	\includegraphics[width=0.32\textwidth]{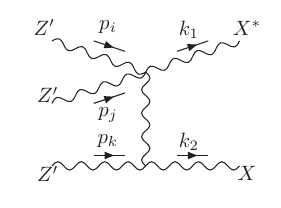}
    	\includegraphics[width=0.32\textwidth]{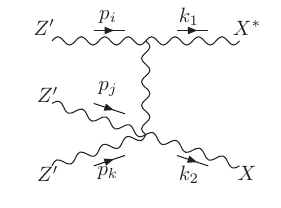}
    	\caption{Feynman diagrams of $Z'+Z'+Z'\to X^\ast+X$ processes. There are $6$ independent diagrams with $i,j,k=1,2,3$ and $i\neq j\neq k$ for the first plot, while $3$ independent diagrams each for the second and third plots.}
    	\label{diagram3to2}
    \end{figure}

    The catalyzed freeze-out production of DM also requires a $Z'+Z'+Z'\to X^\ast+ X$ process whose reaction rate is comparable with the $X^\ast+X\to Z'+Z'$ process during the catalyzed annihilation stage. The diagrams of $Z'+Z'+Z'\to X^\ast+ X$ are shown in FIG.\ref{diagram3to2}. Once the amplitude is written down, the corresponding annihilation cross section can be computed in the non-relativistic limit by using the formula (E4) in Ref.\cite{Cline:2017tka}\footnote{In our definition, the annihilation cross section is $1/S_i$ times of the one defined in Ref.\cite{Cline:2017tka}, where $S_i=n_i\!$ is a symmetry factor from identical initial particles}, and the result is given by
    \beq
    \langle\sigma_3v^2\rangle&\approx&\frac{1}{6}\frac{g_D^6}{192\pi m_{Z'}^5}\left(1-\frac{4}{9}r_{XZ'}\right)^{1/2}f(r_{XZ'}),\nonumber\\
   f(r_{XZ'})&=&\frac{729}{256}r_{XZ'}^{-6}-\frac{243}{16}r_{XZ'}^{-5}+\frac{675}{16}r_{XZ'}^{-4}+\frac{1285}{8}r_{XZ'}^{-3}-\frac{1007}{4}r_{XZ'}^{-2}+\frac{2585}{4}r_{XZ'}^{-1}\nonumber\\
   &&-\frac{2317}{4}+415r_{XZ'}-12r_{XZ'}^2-48r_{XZ'}^3.
    \eeq

    Finally, we need to figure out the decay width of the catalyst. As we have discussed in previous section, catalyst decay due to the dim-5 $G_D$ violated operator \eqref{effop} and thus the decay width must be suppressed by $s_\epsilon^2$. The two-body decay processes are $Z'\to\bar{f}+f,W^++W^-$, where $f$ indicates all type of SM fermions. In the $m_f,m_W\ll m_{Z'}$ limit, the total width can be approximately evaluated as \footnote{The width of $Z'$ in our model is the same as the one given in Ref.\cite{Gabrielli:2015hua}.}
    \beq
    \Gamma_{Z'}\approx\frac{27\alpha t_\epsilon^2 c_\zeta^2 m_{Z'}}{16\hat{c}_W^2}\approx2\times10^{-2}\times t_\epsilon^2 m_{Z'}.
    \eeq
    \subsection{The Boltzmann equations and the solutions}
    The Boltzmann equations of the $X$ and $Z'$ read,
    \beq
    \frac{dn_X}{dt}+3Hn_X&\approx&-\frac{1}{2}\langle\sigma v\rangle_{X^\ast X\to 2\phi_{SM}}(n_X^2-\bar{n}_X^2)\nonumber\\
    &&-\frac{1}{2}\langle\sigma_2 v\rangle\left(n_X^2-\bar{n}_X^2\frac{n_{Z'}^2}{\bar{n}_{Z'}^2}\right)+2\langle\sigma_3 v^2\rangle\left(n_{Z'}^3-\bar{n}_{Z'}^3\frac{n_X^2}{\bar{n}_X^2}\right)\label{Boltzmann_eq1}\\
    \frac{dn_{Z'}}{dt}+3Hn_{Z'}&\approx&-2\langle\sigma v\rangle_{Z'Z'\to 2\phi_{SM}}(n_{Z'}^2-\bar{n}_{Z'}^2)-\langle\Gamma_{Z'}\rangle(n_{Z'}-\bar{n}_{Z'})\nonumber\\
    &&+\frac{1}{2}\langle\sigma_2 v\rangle\left(n_X^2-\bar{n}_X^2\frac{n_{Z'}^2}{\bar{n}_{Z'}^2}\right)-3\langle\sigma_3 v^2\rangle\left(n_{Z'}^3-\bar{n}_{Z'}^3\frac{n_X^2}{\bar{n}_X^2}\right)\label{Boltzmann_eq2}
    \eeq
    where $\bar{n}_i$ is the equilibrium distribution of particle species $i$~\footnote{Note that $n_X$ is defined as the sum of DM and anti-DM densities.}. In a very early era, the Higgs-portal interactions between the dark and SM sectors can thermalize both $X$ and $Z'$. Their number density distributions trace the standard Boltzmann distribution:
    \beq
    n_{X}\approx \bar{n}_X\approx 2\times3\times \left(\frac{m_X T}{2\pi}\right)^{3/2}e^{-m_X/T},\quad n_{Z'}\approx \bar{n}_{Z'}\approx 3\times \left(\frac{m_{Z'} T}{2\pi}\right)^{3/2}e^{-m_{Z'}/T},
    \eeq
    where we have assume that the chemical potentials are negligible. 
    Since there are many parameters involved in the annihilation processes, for simplicity, we parametrize the total annihilation cross sections of $X^\ast+X (Z'+Z')\to\phi_{SM}+\phi_{SM}$ as
    	\beq
    	\langle\sigma v\rangle_{X^\ast X\to 2\phi_{SM}}&=&\frac{g_D^4v_2^2}{256\pi m_X^2v^2}\left(\frac{m_t^2}{m_X^2}\right)\xi_{X}^2,\\
    	\langle\sigma v\rangle_{Z'Z'\to 2\phi_{SM}}&=&\frac{g_D^4v_2^2}{512\pi m_{Z'}^2v^2}\left(\frac{m_t^2}{m_{Z'}^2}\right)\xi_{Z'}^2,
    	\eeq
    	where $\xi_X$ and $\xi_{Z'}$ encode the effects of all the Higgs-portal channels and relevant parameters, e.g. $\alpha$, $\alpha_{13}$, $\alpha_{23}$.
    As the temperature decreases, these processes fall behind the Hubble expansion and we assume that it happens before $T\sim m_X/10$. After that, the first terms in the right-handed sides of Boltzmann equations for both $X$ and $Z'$ can be dropped. At the moment, let us assume that the decay and inverse decay terms of $Z'$ are negligible before the DM freeze-out and thus the second term in the right-handed side of the Boltzmann equation for $Z'$ can be dropped too.

    DM and catalyst still keep in thermal equilibrium since both $X^\ast+X\leftrightarrow Z'+Z'$ and $Z'+Z'+Z'\leftrightarrow X^\ast+X$ reactions are efficient enough for forcing the distributions to satisfy
    \beq
    \frac{n_X}{\bar{n}_X}\approx\frac{n_{Z'}}{\bar{n}_{Z'}},\qquad \left(\frac{n_X}{\bar{n}_X}\right)^2\approx\left(\frac{n_{Z'}}{\bar{n}_{Z'}}\right)^3~.
    \eeq
    The only reasonable solutions to these equations are $n_X=\bar{n}_{X},~ n_{Z'}=\bar{n}_{Z'}$.

\begin{figure}[tb]
    	\centering
    	\includegraphics[width=0.45\textwidth]{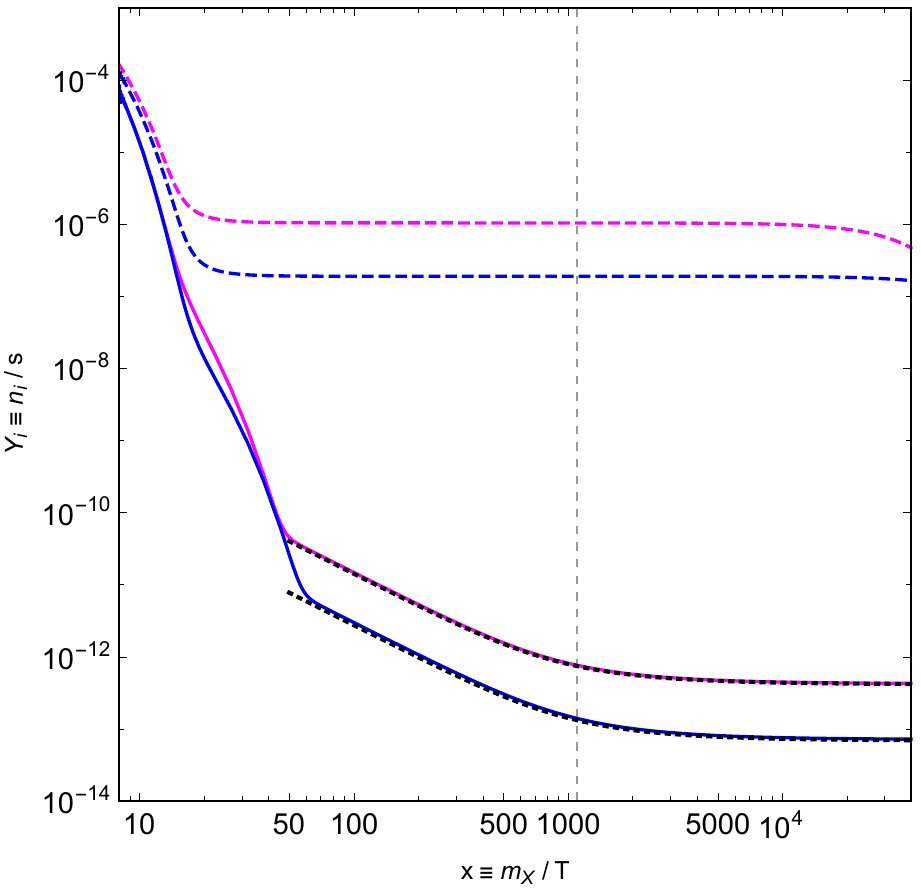}
    	\includegraphics[width=0.457\textwidth]{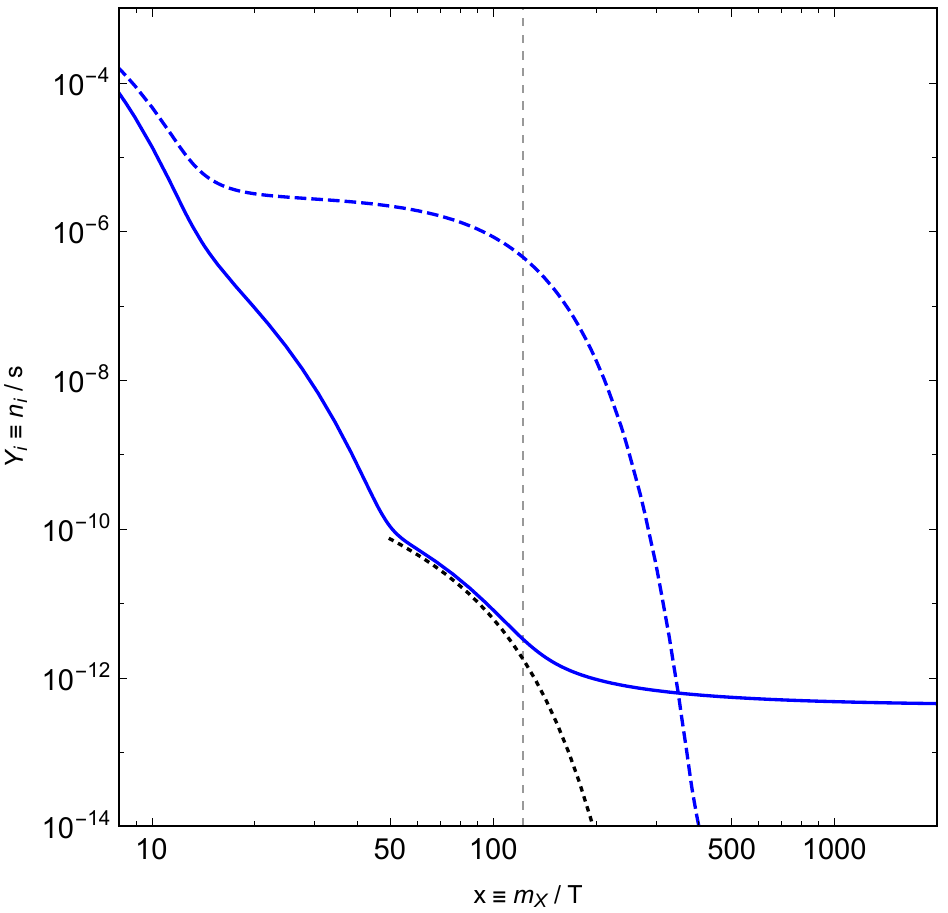}
    	\caption{The evolutions of $Y_{Z'}(x)$ (dashed lines) and $Y_{X}(x)$ (solid lines). In the left panel, $t_\epsilon=10^{-11},~\xi_X=\xi_{Z'}=10^{-5}$ is chosen for the three benchmark models with $(m_X,r_{XZ'}^{1/2},g_D)=(1~\textrm{TeV},1.32,1.015)$ (magenta lines), and $(6~\textrm{TeV},1.25,2.55)$ (blue lines). The black dotted lines are the approximated results of $Y_X(x)$ given by eq.\eqref{apprsolution}. In the right panel, $m_X=1~\textrm{TeV},~r_{XZ'}^{1/2}=1.3,~g_D=0.68,~t_\epsilon=5\times10^{-9},~\xi_X=\xi_{Z'}=10^{-5}$ is chosen. The black dotted line is the approximate solution of $Y_X(x)$ given by eq.\eqref{apprsolution2}.}
    	\label{benchmark}
    \end{figure}

    To figure out the temperature of departure from chemical equilibrium, we can firstly sum up the two equations, \eqref{Boltzmann_eq1} and \eqref{Boltzmann_eq2}, and get
    \beq
    \frac{d(n_{Z'}+n_X)}{dt}+3H(n_{Z'}+n_X)\approx -\langle\sigma_3 v^2\rangle\left(n_{Z'}^3-\bar{n}_{Z'}^3\frac{n_X^2}{\bar{n}_X^2}\right).
    \eeq
    Since $X$ is heavier than $Z'$ and thus $n_X\approx\bar{n}_X\ll \bar{n}_{Z'}$, we can neglect the $n_X$ in the left-handed side of the equation. The evolution of $n_{Z'}$ is now determined only by the competition between $\langle\sigma_3v^2\rangle$ term and the Hubble expansion. We can expect that $Z'$ freezes out when
    \beq
    \langle\sigma_3 v^2\rangle n_{Z'}^3\simeq Hn_{Z'},
    \eeq
    To determine the departure temperature $T_c$ (or $x_c\equiv m_X/T_c$) more precisely, we define  $n_{Z'}\equiv \bar{n}_{Z'}(1+\delta(x))$ and introduce $x\equiv m_X/T,~Y_{Z'}\equiv n_{Z'}/s$, where $s=(2\pi^2/45)g_\ast T^3$ is the entropy density\footnote{In this work, since the temperature we consider is above several GeV, we make an approximation that $g_{\ast s}\approx g_\ast$ for simplicity.}.
    Note that $\langle\sigma_2v^2\rangle n_{X}^2$ is still much larger than $Hn_X$ at $T=T_c$, so $n_X$ is forced to satisfy $n_X/\bar{n}_X=n_{Z'}/\bar{n}_{Z'}$. Using the relation $Y_X/\bar{Y}_X=Y_{Z'}/\bar{Y}_{Z'}=1+\delta$, the Boltzmann equation of $Z'$ becomes
    \beq
    \frac{d\ln\bar{Y}_{Z'}}{dx}(1+\delta)+\frac{d\delta}{dx}\approx-\frac{\lambda_X}{x^2}\langle\sigma_3v^2\rangle s\bar{Y}_{Z'}^2(1+\delta)^2\delta,
    \eeq
    where $\lambda_X\equiv \sqrt{\pi g_\ast/45} m_Xm_{pl}$.
    Since $Y_{Z'}$ closely trace the equilibrium distribution, $d\delta/dx$ term is negligible before $Y_{Z'}$ frozen. We can take a reference quantity $\delta_c\equiv\delta(x_c)\sim2.5$ as a sign of $Z'$ starting departure from thermal equilibrium, then $x_c$ can be approximately determined by
    \beq\label{solxc}
    x_c=r_{XZ'}^{1/2} W_0(\sqrt{A}),\quad A\equiv\frac{9\lambda_X\langle\sigma_3v^2\rangle m_X^3(1+\delta_c)\delta_c}{(2\pi)^5r_{XZ'}^2\left(1-\frac{3r_{XZ'}^{1/2}}{2x_a}\right)}\left(\frac{90}{g_\ast}\right),
    \eeq
    where $x_a\sim16$ is an approximate value of $x_c$ coming from the first iteration of \eqref{solxc} and $W_0(z)$ is the principle branch of Lambert $W$ function.
    After $T\gtrsim T_c$, $n_{Z'}$ starts to deviate from the Boltzmann suppressed equilibrium distribution, and thus $Y_{Z'}>\bar{Y}_{Z'}$. The equation of $Y_{Z'}$ can be approximated with
    \beq\label{BEZpfo}
    \frac{dY_{Z'}}{dx}\approx -\frac{\lambda_X}{x^5}\langle\sigma_3v^2\rangle (2\pi)^2m_X^3\left(\frac{g_\ast}{90}\right)Y_{Z'}^3~.
    \eeq
    An approximate solution of eq.\eqref{BEZpfo} in $x>x_c$ is given by
    \beq\label{frYZp}
    Y_{Z'}(x)\approx\frac{\bar{Y}_{Z'}(x_c)(1+\delta_c)}{\sqrt{1+\frac{x_c(1+\delta_c)}{2\delta_c}\left(r_{XZ'}^{-1/2}-\frac{3}{2x_a}\right)\left(1-\frac{x_c^4}{x^4}\right)}}.
    \eeq
    We can see that $Y_{Z'}(x)$ quickly tends to a fixed quantity after $x>x_c$.
    After $Z'$ freezes out, the process $X^\ast+X\leftrightarrow Z'+Z'$ is still efficient and thus the DM and catalyst are in chemical equilibrium. The distribution of $X$ can be determined by
    \beq
    Y_X(x)\approx \frac{\bar{Y}_X}{\bar{Y}_{Z'}}Y_{Z'}\approx 2r_{XZ'}^{3/4}e^{-(1-r_{XZ'}^{-1/2})x}Y_{Z'}(x)
    \eeq
    Since $m_{Z'}<m_{X}$, the reaction rate of process $Z'+Z'\to X^\ast+X$ is exponentially decreasing and it finally fades out. To be precise, this happens when
    	\beq
    	\frac{1}{2}\langle\sigma_2 v\rangle \frac{\bar{n}_X^2}{\bar{n}_{Z'}^2}n_{Z'}^2\lesssim 2\langle\sigma_3 v^2\rangle n_{Z'}^3~\Rightarrow~ e^{-2(1-r_{XZ'}^{-1/2})x}\lesssim\left[\frac{4\langle\sigma_3 v^2\rangle}{\langle\sigma_2 v\rangle}m_X^3\left(\frac{2\pi^2g_\ast}{45}\right)Y_{Z'}\right]x^{-3}~.\nonumber
    	\eeq
    The exponential function in the left-handed side decreases much faster than $x^{-3}$ in the right-handed side when $x$ grows, and thus $3Z'\to X^\ast+X$ becomes more efficient than $2Z'\to X^\ast+X$ in lower temperature, so that we can neglect the $2Z'\to X^\ast+X$ processes in the catalyzed annihilation era. After that, the equation of $Y_X$ becomes
    \beq\label{BeqYX}
    \frac{dY_X}{dx}\approx \frac{\lambda_X}{x^2}\left[-\frac{1}{2}\langle\sigma_2 v\rangle Y_X^2+2\langle\sigma_3 v^2\rangle sY_{Z'}^3\right]
    \eeq
    The DM depletes through $X^\ast+X\to Z'+Z'$ and $Z'+Z'+Z'\to X^\ast+X$ processes which means the catalyzed annihilation stage starts. $Y_X$ in this era is given by
    \beq
    \tilde{Y}_X(x)= C_Xx^{-3/2},
    \eeq
    where
    \beq\label{CX}
    C_X\equiv4\pi\left(\frac{g_\ast}{90}\right)^{1/2}\left(\frac{\langle\sigma_3v^2\rangle}{\langle\sigma_2v\rangle}\right)^{1/2}m_X^{3/2}Y_{Z'}^{3/2}~.
    \eeq
    The catalyzed annihilation stage ends when
    \beq
    \langle\sigma_2 v\rangle n_X^2\simeq\langle\sigma_3 v\rangle n_{Z'}^3\simeq Hn_X~,
    \eeq
    and then DM freezes out. There is a good approximate solution of eq.\eqref{BeqYX}:
    \beq\label{apprsolution}
    Y_X(x)\approx \tilde{Y}_X(x)f_X(z)\quad,\quad f_X(z)\equiv\frac{K_{\frac{4}{5}}(z)}{{K_{\frac{1}{5}}(z)}}
    \eeq
    with $z\equiv\frac{2A_X}{5x^{5/2}}$, where $A_X$ is defined by
    \beq
    A_X\equiv\frac{1}{2}\lambda_X\langle\sigma_2v\rangle C_X\quad,
    \eeq
    $K_\alpha(z)$ is the modified Bessel function of the second kind.
    We can check that in the large $z$ limit $f_X(z)\to1$, while in the small $z$ limit $f_X(z)\to[\Gamma(4/5)/\Gamma(1/5)](z/2)^{-3/5}\propto x^{3/2}$. Therefore, $Y_X(x)$ traces $\tilde{Y}_X(x)$ before DM freezes out ($z\gg1$) and approaches a constant after freeze out ($z\ll1$). We define $Y_X^{fo.}$ to denote the final value of $Y_X(x)$ after DM freezes out and its explicit expression is given by
    \beq
    Y_X^{fo.}=\frac{\Gamma(4/5)}{\Gamma(1/5)}\left(\frac{A_X}{5}\right)^{-3/5}\quad.
    \eeq
    Finally, the relic abundance of DM today can be computed by
    \beq
    \Omega_X h^2=2.83\times10^{11}\times\left(\frac{m_X}{1~\mathrm{TeV}}\right)Y_X^{fo.}.
    \eeq
    We solve the Boltzmann equations numerically for two different benchmark models:
    \begin{enumerate}
    	\item $m_X=1$~TeV, $r_{XZ'}^{1/2}=m_X/m_{Z'}=1.32$, $g_D=1.015$, $t_\epsilon=10^{-11}$, $\xi_X=\xi_{Z'}=10^{-5}$ (magenta lines of left panel in FIG.\ref{benchmark}),
    	\item $m_X=6$~TeV, $r_{XZ'}^{1/2}=m_X/m_{Z'}=1.25$, $g_D=2.55$, $t_\epsilon=10^{-11}$, $\xi_X=\xi_{Z'}=10^{-5}$ (blue lines of left panel in FIG.\ref{benchmark}),
  	\end{enumerate}
  which can reproduce the observed relic abundance of DM $\Omega_X h^2\approx 0.12$~\cite{Aghanim:2018eyx}.
  The evolution of the $Y_{Z'}(x)$ and $Y_X(x)$ are shown in the left panel of FIG.\ref{benchmark}~. Solid lines represent $Y_{X}(x)$, while the dashed lines represent $Y_{Z'}(x)$. The black dotted lines represent the analytical approximate solutions of $Y_X(x)$ given by eq.\eqref{apprsolution}. We find that our approximate results match the numerical ones very well. The temperature of $Z'$ freezing is around $T_c\approx m_X/16$. The temperature of DM freeze-out is about $T_f\approx m_X/10^3$ (vertical dashed line in FIG.\ref{benchmark}) given by $z\approx0.3$ ( where $Y_X(x)$ is about $1.6$ times of $\tilde{Y}_X(x)$).

    Now we can determine the constraint on the decay width of the catalyst. The condition is
    \beq
    \langle\Gamma_{Z'}\rangle \ll H(T_f)\quad\Rightarrow\quad t_\epsilon \ll2\times10^{-10},
    \eeq
    for the three chosen benchmark models. In the case with $m_X=1$~TeV, current direct detection bound on the magnetic moment of DM is about~\cite{Hisano:2020qkq}
    \beq
    \left|\frac{\mu_X}{\mu_N}\right|\lesssim 10^{-5},
    \eeq
    where $\mu_N=e/2m_p$ is the proton magnetic moment. The dark matter magnetic moment can be estimated by $\mu_X\sim (e/2m_X)(g_D\hat{c}_Ws_\epsilon/2)$ and thus the bound on the $s_\epsilon\approx t_\epsilon$ is about
    \beq
    s_\epsilon\lesssim 0.05~,
    \eeq
    which is much looser than the constraint from decay width.

\begin{figure}[tb]
    	\centering
    	\includegraphics[width=0.6\textwidth]{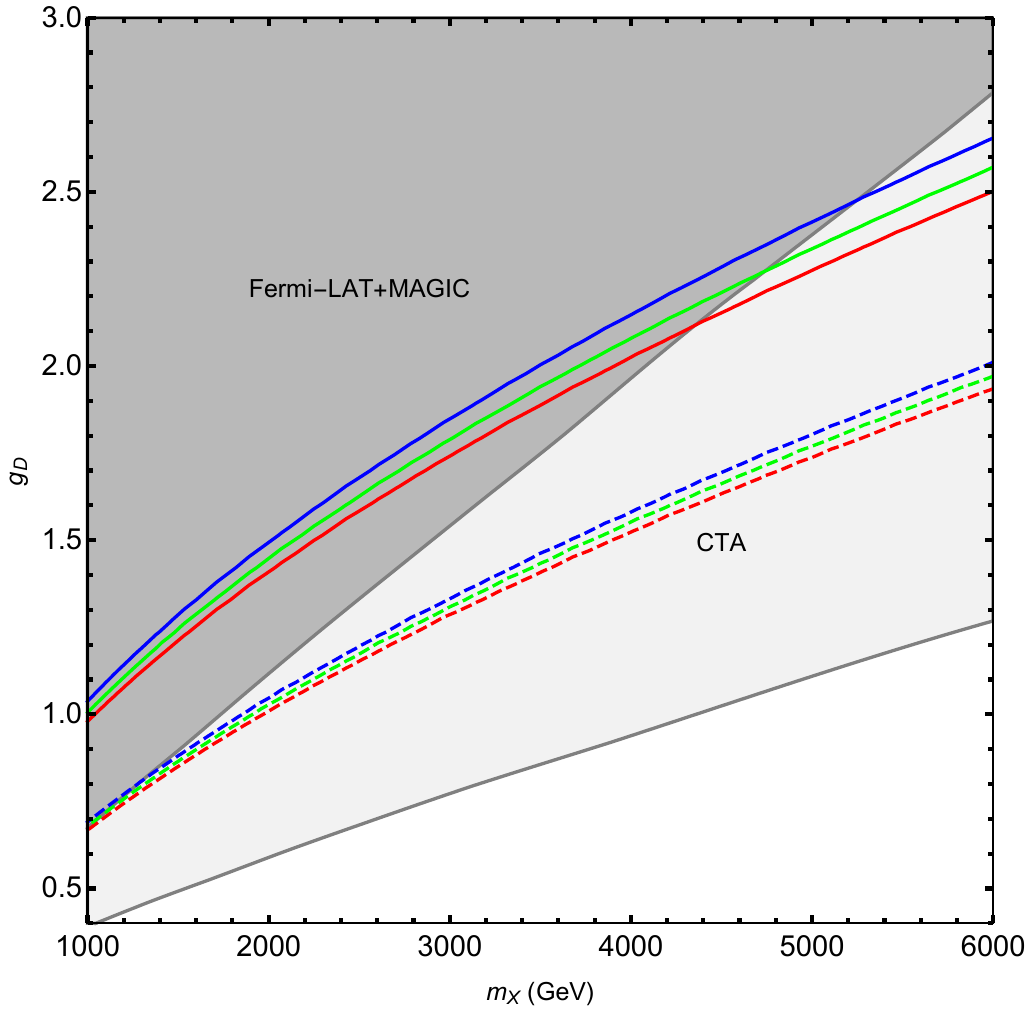}
    	\caption{The dark gray region is excluded by Fermi-LAT data~\cite{Hoof:2018hyn} at 95\% CL, while the light gray region is the prospect of CTA experiment~\cite{Doro:2012xx}. Solid lines correspond to the parameters that reproduce $\Omega_Xh^2=0.12$~\cite{Aghanim:2018eyx} by choosing $r_{XZ'}^{1/2}=1.2$ (red), $1.3$ (green), and $1.4$ (blue) and fixing $t_\epsilon=10^{-11},~\xi_X=\xi_{Z'}=10^{-5}$. Dashed lines represent the cases by fixing $t_\epsilon=5\times10^{-9},~\xi_X=\xi_{Z'}=10^{-5}$.}
    	\label{IDconstraint}
    \end{figure}

    Although the model is unlikely to be constrained by the DM direct detection experiment, it can have significant signal in the indirect detection experiments. For example, remnant of DM in dwarfs satellite galaxies can annihilate each other and produce catalysts, and then catalysts will decay into SM particles. These processes can contribute to the continuous spectrum of $\gamma$-ray and then be probed by the Fermi-LAT experiments~\cite{Hoof:2018hyn}. The absence of signals put stringent constraints on the parameter space of the models. In FIG.\ref{IDconstraint}, the dark gray region has been excluded by current Fermi-LAT data, while the light gray region is an estimation of future CTA experiment sensitivity. The solid colored lines represents the parameters which can obtain the $\Omega_Xh^2=0.12$ for $r_{XZ'}^{1/2}=1.2$ (red), $1.3$ (green), and $1.4$ (blue) with fixing $t_\epsilon=10^{-11},~\xi_X=\xi_{Z'}=10^{-5}$. We find that the region $m_X\lesssim4.5$~TeV has been excluded by the Fermi-LAT observation at 95\% CL. The whole region of our interest is covered by the prospects of CTA sensitivity~\cite{Doro:2012xx}, so our model can be tested in the next generation of high energy $\gamma$-ray observation.

    Note that in the above discussions, we have assumed that the catalyst particle decay after the DM freezes out. When $t_\epsilon>10^{-10}$, this assumption is not valid anymore. Consider the case that $Z'$ is long-living enough for starting the catalyzed annihilation but it decays before $X$ freezes out. The equation of $Y_{Z'}$ becomes
    \beq
    \frac{dY_{Z'}}{dx}\approx -\lambda_X\frac{\Gamma_{Z'}}{(2\pi)^2m_X^3}\left(\frac{90}{g_\ast}\right)x(Y_{Z'}-\bar{Y}_{Z'})~.
    \eeq
    which has an approximate solution of the form
    \beq
    Y_{Z'}(x)\approx \tilde{Y}_{Z'} e^{-\frac{C_{Z'}}{2}x^2}
    \eeq
    where $ \tilde{Y}_{Z'} $ is the $x\to\infty$ limit of \eqref{frYZp}, and $C_{Z'}\equiv(90\lambda_X\Gamma_{Z'})/((2\pi)^2g_\ast m_X^3)$. We can see that $Y_{Z'}(x)$ starts to fastly decrease when $C_{Z'}x^2\sim1$. At the same time, $Y_X(x)$ in the catalyzed annihilation era should be
    \beq\label{apprsolution2}
     \hat{Y}_X(x)= C_Xx^{-3/2}e^{-\frac{3C_{Z'}}{4}x^2}~.
    \eeq
    The freeze-out of $X$ happens when
    \beq
    x_f\approx \left(\frac{3}{C_{Z'}}W_0\left[\left(\frac{2\delta_f(2+\delta_f)A_X}{3(1+\delta_f)C_{Z'}}\right)^{4/9}\frac{C_{Z'}}{3}\right]\right)^{1/2}~.
    \eeq
    The approximate result of $Y_X(\infty)$ after $x>x_f$ is given by
    \beq
    Y_X(\infty)\approx \frac{\hat{Y}_X(x_f)(1+\delta_f)}{1+\frac{3(1+\delta_f^2)}{2(2+\delta_f)\delta_f}(1+C_{Z'}x_f^2)}~,
    \eeq
    where $\delta_f=1.3$ can reproduce the numerical result well.
    In the right panel of FIG.\ref{benchmark}, we show the evolution of $Y_X(x)$ (blue solid) and $Y_{Z'}(x)$ (blue dashed) from numerical computation for a benchmark model with $m_X=1~\textrm{TeV},~r_{XZ'}^{1/2}=1.3,~g_D=0.68,~t_\epsilon=5\times10^{-9},~\xi_X=\xi_{Z'}=10^{-5}$. The black dotted line is the approximate solution of $Y_X(x)$ before DM freezes out. We can see that the freeze-out of $X$ is triggered by the decay of $Z'$, therefore the freeze-out temperature also depends on the decay width of $Z'$. In the FIG.\ref{IDconstraint}, we show the dashed lines representing the parameters achieving the observed DM relic abundance by choosing $t_\epsilon=5\times 10^{-9},~\xi_X=\xi_{Z'}=10^{-5}$. The red, green, blue colors corresponds to $r_{XZ'}^{1/2}=1.2,~1.3,~1.4$. Since $g_D$ for reproducing the DM relic abundance is smaller in this case, the region with $m_X>1$~TeV survives from the Fermi-LAT bound. The CTA sensitivity also covers all the dashed lines of the model, so we can expect our models to be tested in the future experiments.
    
    \section{Kinetic equilibrium before DM freeze-out}\label{sect4}
     As pointed out in Ref.~\cite{Xing:2021pkb}, it is not easy to keep DM in kinetic equilibrium (KE) with the thermal bath at a temperature as low as $T_f\sim m_X/10^3$. The reason is that the couplings corresponding to DM annihilation are usually the same as the ones leading to scattering. If the catalyst particle is required to decouple early ($T>T_c\sim m_X/16$), the corresponding coupling should be very small. On the other hand, a tiny coupling also suppresses the reaction rate of scattering processes, and thus the scattering rate may quickly fall behind the Hubble expansion rate. In previous sections, the annihilation processes of DM to SM particles are mediated by Higgs bosons, whose couplings are set to be as small as $\xi_{X}\sim\xi_{Z'}\lesssim10^{-4}$, and thus the scattering rate of DM with the thermal bath is also extremely suppressed in late time.

     In the following discussion, we will discuss two ways for keeping DM in KE until freeze-out. One way is to consider a stronger Higgs-portal coupling, and the other way is to couple the DM with a thermal Axion-Like Particle (ALP) $\eta$ during the catalyzed annihilation era. 
     
     \subsection{ The Higgs-portal strategy}
     If the Higgs-portal interaction among the dark and SM sector becomes stronger, it is possible to keep the DM particle in KE until freeze-out. The cross section of scattering between $X$ and SM fermion $f$, through the Higgs-portal, is given by
      \beq
      \langle\sigma v\rangle_{Xf\to Xf}&\approx& \frac{c_fg_{D}^4v_2^2m_f^2}{384\pi m_X^4v^2}\left[(c_\alpha\alpha_{13}+s_\alpha\alpha_{23})+\frac{4v_3}{v_2}(c_\alpha\alpha_{23}-s_\alpha\alpha_{13})\right]^2x^3\mathcal{F}(x)~,\\
      \mathcal{F}(x)&=&\int_0^\infty dq e^{-xq}\mathcal{I}(x)~,\nonumber\\
      \mathcal{I}(x)&\equiv&\frac{q^2}{4(1+2q)^2(4q^2+2r_{hX}q+r_{hX})}\nonumber\\
      &&\quad\times[8q^4+(32-r_{hX})q^3-(32-26r_{hX}+12r_{hX}^2)q^2\nonumber\\
      &&\qquad-(48-32r_{hX}+12r_{hX}^2)q-(12-8r_{hX}+3r_{hX}^2)]\nonumber\\
      &&+\frac{12-8r_{hX}+3r_{hX}^2}{16}\ln\left(1+\frac{4q^2}{r_{hX}(1+2q)}\right)~,
      \eeq
      where $c_f$ is the color number of SM fermion, $f$, while $r_{hX}\equiv m_1^2/m_X^2$. 
      The KE condition for DM particle can be estimated by~\cite{Hofmann:2001bi,Visinelli:2015eka}
      \beq
      \left(\frac{T}{m_X}\right)\frac{\Gamma_{el}}{H}\gtrsim 1~,\qquad \Gamma_{el}\approx\sum_{f=\tau,c,b} n_f\langle\sigma v\rangle_{Xf\to Xf}~.
      \eeq
      where $H$ is the Hubble expansion rate. Note that a suppression factor $(T/m_X)$ is introduced since a non-relativistic particle requires multiple scatterings with the plasma for transferring its momentum.
       
      Since there are lots of parameters in the potential, we only focus on two special cases. The first case is to consider that $\lambda_{02}=0,~s_\alpha\ll1$, and thus the annihilation cross sections of $Z'+Z'\to \phi_{SM}+\phi_{SM}$ are suppressed by $s_\alpha^2$ (see eq.\eqref{alpha13},~\eqref{alpha23}, and \eqref{anntoSM}). On the other hand, the scatterings between the DM particle and SM fermions are not suppressed by tiny $s_\alpha$ since $c_\alpha\alpha_{23}-s_\alpha\alpha_{13}\approx -\lambda_{03}v_3v/m_2^2$. Therefore, we can always find a $\lambda_{03}$ which is large enough for maintaining the KE of DM during the catalyzed annihilation era, without affecting the freeze-out of $Z'$ in the early era ($x_c\sim16$). 
      
\begin{figure}[tb]
  	\centering
  	\includegraphics[width=0.45\textwidth]{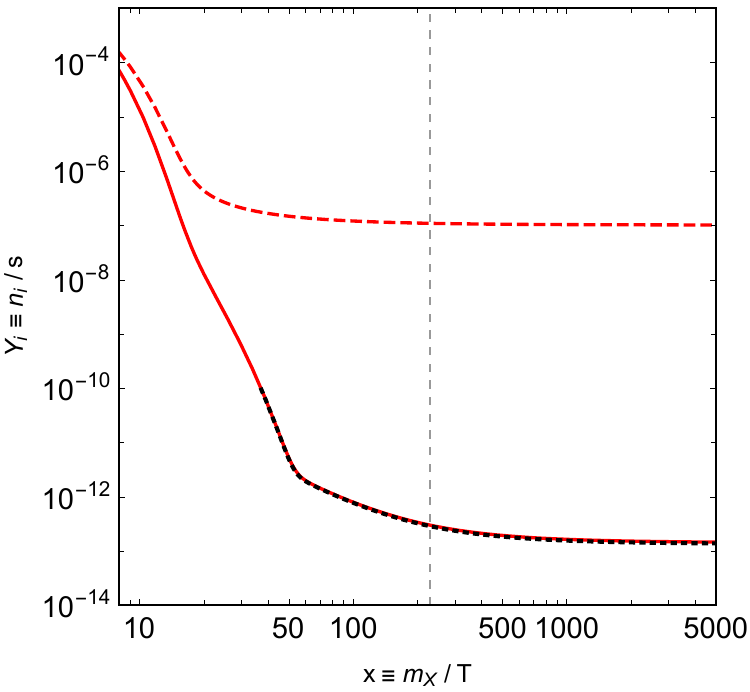}
  	\includegraphics[width=0.45\textwidth]{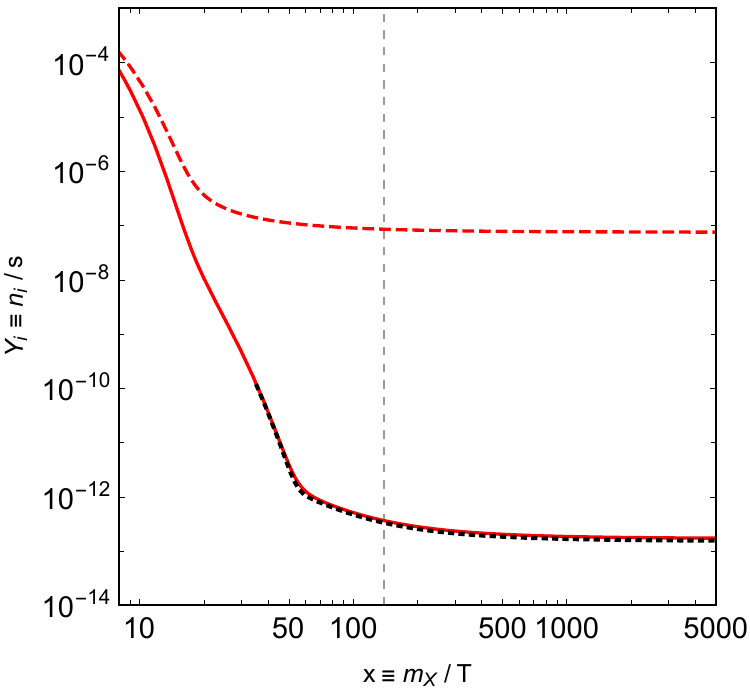}
  	\caption{The evolutions of $Y_{Z'}(x)$ (red dashed line) and $Y_{X}(x)$ (red solid line). The black dotted line is given by the approximate solution. The vertical gray dashed lines indicate the $x_f$ for DM particle. Left panel: Higgs-portal model with parameters, $m_X=3~\textrm{TeV}$,~$r_{XZ'}^{1/2}=1.3$,~$g_D=1.24$,~$t_\epsilon=10^{-11}$,~$\xi_{eff}=1.5\times10^{-3}$. Right panel: ALP-portal model with parameters, $m_X=2.5~\textrm{TeV}$,~$r_{XZ'}^{1/2}=1.3$,~$g_D=1.0$,~$m_X/\Lambda'=0.036$,~$t_\epsilon=10^{-11}$,~$\xi_X=\xi_{Z'}=10^{-5}$. }
  	\label{benchmark2}
  \end{figure}

      For another special choice of parameter that, $\lambda_{03}=0,~s_\alpha\ll1$, there is a common Higgs-portal coupling for $X$ and $Z'$ since $c_\alpha\alpha_{23}-s_\alpha\alpha_{13}\approx s_\alpha$ is negligible, so that we can simplify the analysis by defining an effective coupling 
      \beq
      \xi_{eff}=\left(\frac{v_2}{v}\right)\left(c_\alpha\alpha_{13}+s_\alpha\alpha_{23}\right)~.
      \eeq
      The annihilation cross sections for $Z'+Z'\to\phi_{SM}+\phi_{SM}$ and $X+f\to X+f$ become
      \beq
      \langle\sigma v\rangle_{Z'Z'\to 2\phi_{SM}}&\approx&\frac{g_D^4\xi_{eff}^2}{256\pi m_{Z'}^2}\left(\frac{m_t^2}{m_{Z'}^2}+\frac{4}{3}\right)~,\\
      \langle\sigma v\rangle_{Xf\to Xf}&\approx& \frac{g_{D}^4\xi_{eff}^2c_fm_f^2}{384\pi m_X^4}\cdot x^3\mathcal{F}(x)~.
      \eeq
      Since a larger $\xi_{eff}$ can give rise to a more efficient annihilation rate for the catalyst particles, it is possible that the freeze-out temperature of $Z'$ is also determined by 2 to 2 annihilation, $Z'+Z'\to \phi_{SM}+\phi_{SM}$, rather than only by $Z'+Z'+Z'\to X^\ast +X$. These additional annihilation channels also accelerate the freeze-out of DM since $\tilde{Y}_{Z'}$ is reduced. In the left panel of FIG.\ref{benchmark2}, we show the evolution of $Y_{X,Z'}$ for a benchmark model with parameters: $m_X=3~\textrm{TeV},~r_{XZ'}^{1/2}=1.3,~g_D=1.24,~t_\epsilon=10^{-11},~\xi_{eff}=1.5\times10^{-3}$. The freeze-out temperature of DM is found to be $x_f\approx 230\Rightarrow T_f\approx13$~GeV.
    
    In the left panel of FIG.\ref{IDconstraint2}, the indirect detection, KE bound\footnote{Since the KE bound also depends on $\xi_{eff}$, its limit line (dashed brown) is obtained by connecting the bound points for different $\xi_{eff}$.}, and five contours reproducing $\Omega_Xh^2=0.12$ are shown in the $m_X-g_D$ plane. Five colored lines represent models with $\xi_{eff}=10^{-4}$ (red), $5\times10^{-4}$ (magenta), $10^{-3}$ (green), $2\times10^{-3}$ (blue), $4\times10^{-3}$ (purple), by fixing $r_{XZ'}^{1/2}=1.3,~t_\epsilon=10^{-11}$. We find that for $\xi_{eff}\gtrsim10^{-3}$, Higgs-portal scattering allows DM to freeze out before out of KE in our interested parameter region. For $\xi_{eff}>2\times10^{-3}$, DM can be as light as $m_X=1.4$~TeV. 
    
    Note that for the case with $\xi_{eff}\lesssim10^{-3}$, it is also possible to maintain the KE of $X$ before freeze-out if the catalyst $Z'$ decays earlier (but it should be guaranteed to not decay before the catalyzed annihilation starts). This can be understood since the decay of $Z'$ leads to a sudden freeze-out of $X$, so that $X$ can easily freezes out before kinetic decoupling.

     \subsection{The ALP strategy}
     Another way to postpone the kinetic decoupling of DM is considering an axion-like particle (ALP) extension. The ALP couples to the SM and the dark sector in the following form
    \beq
    \mathcal{L}_\eta\supset -\frac{\eta}{\Lambda'} \tilde{V}^{a}_{\mu\nu} V^{a,\mu\nu}-C_f\sum_{f=q,l,\nu}\frac{1}{\Lambda'}(\partial_\mu \eta) \bar{f}\gamma^\mu\gamma^5f+...~,
    \eeq
    where we have simplified our model by assuming that $C_f$ is common for all species of SM fermions, and the coupling between $\eta$ and SM gauge bosons are negligible. If the ALP has a mass around $1$~GeV~\footnote{An ALP with a mass $\sim1$~GeV and $\Lambda'\gtrsim 30$~TeV is consistent with most of current experimental constraints~\cite{Bauer:2017ris}.}, it can easily keep in thermal equilibrium due to its decay and inverse decay. The cross sections of scattering processes $X+\eta\to X+\eta$ and $X+f\to X+f$ can be derived as
    \beq
    \langle\sigma v\rangle_{X\eta}&\approx& \frac{4m_X^2}{27\pi\Lambda'^4}x^{-2}~,\\
    \langle\sigma v\rangle_{Xf}&\approx& \frac{8c_f(C_f)^2m_f^2}{3\pi \Lambda'^2}=\langle\sigma v\rangle_{X\eta}\times18c_f(C_f)^2\left(\frac{m_f}{T}\right)^2~,
    \eeq
    where $c_f$ is the color number of SM fermion, $f$. Taking all dominant fermion species, $\tau,~c$ and $b$ into account, the collision rate of $X+f\to X+f$ scattering is 
    \beq
   \Gamma_{Xf}= \sum_{f=\tau,c,b}n_f\langle\sigma v\rangle_{Xf}\approx \Gamma_{X\eta}C_f^2\left(\frac{47~\textrm{GeV}}{T}\right)^2,\qquad (T\gtrsim m_b)
    \eeq
    where $\Gamma_{X\eta}=n_\eta \langle\sigma v\rangle_{X\eta}$ is the collision rate of $X-\eta$ scattering. If we take $C_f=1$, we can see that the $X+\eta\to X+\eta$ scattering dominates in a temperature $T\gtrsim50$~GeV, while the $X+f\to X+f$ scatterings dominate when the temperature drops below $50$~GeV. 
   The requirement that kinetic equilibrium is maintained until DM freeze-out leads to a condition:
    \beq
    \left(\frac{T}{m_X}\right)\frac{\Gamma_{X\eta}+\Gamma_{Xf}}{H}\gtrsim1~,
    \eeq
    at $T=T_f=m_X/x_f$.

\begin{figure}[tb]
	\centering
	\includegraphics[width=0.45\textwidth]{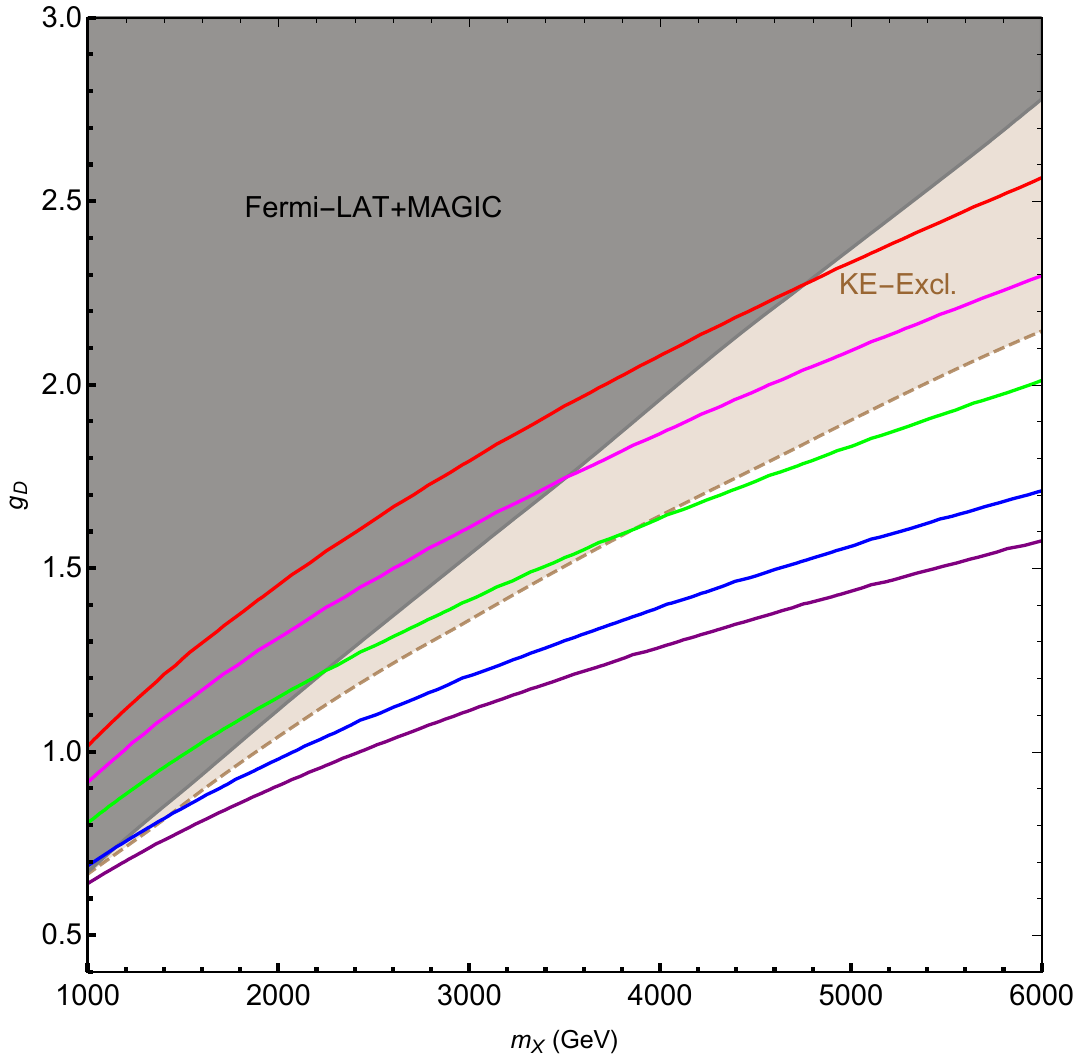}
	\includegraphics[width=0.45\textwidth]{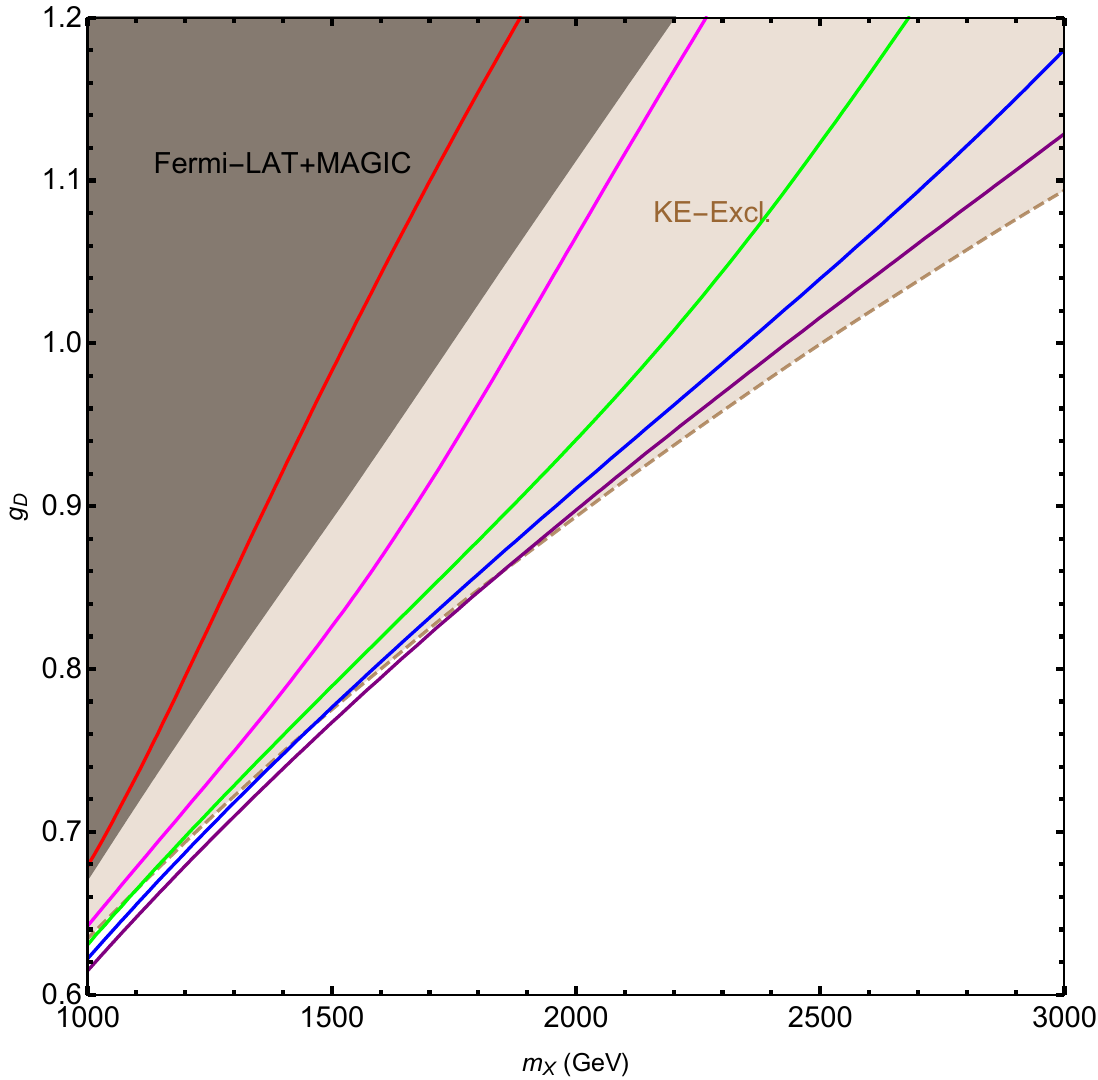}
	\caption{The dark gray region is excluded by Fermi-LAT data~\cite{Hoof:2018hyn} at 95\% CL, while the brown region is excluded by out of kinetic equilibrium of DM. Solid lines correspond to the parameters which can reproduce $\Omega_Xh^2=0.12$~\cite{Aghanim:2018eyx}. Left panel: colored lines indicate $\xi_{eff}=10^{-4}$~(red), $5\times10^{-4}$~(magenta), $10^{-3}$~(green), $2\times10^{-3}$~(blue), $4\times10^{-3}$~(purple) by fixing $r_{XZ'}^{1/2}=1.3,~t_\epsilon=10^{-11}$. Right panel: colored lines indicate  $m_X/\Lambda'=0.020$~(red),~$0.025$~(magenta),~$0.030$~(green),~$0.035$~(blue),~$0.040$~(purple) by fixing $r_{XZ'}^{1/2}=1.3,~t_\epsilon=10^{-11},~\xi_X=\xi_{Z'}=10^{-5}$.}
	\label{IDconstraint2}
\end{figure}

    On the other hand, the annihilation rates of $Z'+Z'\to \eta +\eta$ and $Z'+Z'\to\eta\to \bar{t}+t$ at $x_c$ are given by
    \beq
    \langle\sigma v\rangle_{2Z'\to2\eta}\bar{n}_{Z'}&\approx& \frac{r_{XZ'}^{-1}}{18\pi m_X^2}\left(\frac{m_X}{\Lambda'}\right)^4\times3\times \frac{m_X^3}{(2\pi)^{3/2}}x_c^{-3/2}e^{-x_c}~,\\
    \langle\sigma v\rangle_{2Z'\to\bar{t}t}\bar{n}_{Z'}&\approx& \frac{4C_f^2m_t^2}{\pi\Lambda'^4}x_c^{-1}\times3\times \frac{m_X^3}{(2\pi)^{3/2}}x_c^{-3/2}e^{-x_c}~,
    \eeq
    which are usually larger than the rate of $Z'+Z'+Z'\to X^\ast +X$. It means that the freeze-out temperature of $Z'$ is determined by the effective ALP scale, $\Lambda'$, rather than by the gauge couping, $g_D$. Since $Y_{Z'}$ directly affect the evolution of $Y_X$ during the catalyzed annihilation era, the final relic density of DM also depends on $m_X/\Lambda'$. In the right panel of FIG.\ref{benchmark2}, we show the evolutions of $Y_X$ and $Y_{Z'}$ for a benchmark model with parameters: $m_X=2.5~\textrm{TeV}$, $r_{XZ'}^{1/2}=1.3$, $g_D=1.0$, $m_X/\Lambda'=0.036$, $t_\epsilon=10^{-11}$, $\xi_X=\xi_{Z'}=10^{-5}$. We find that $Z'$ freeze-out at $x_c\approx20$ and the catalyzed annihilation of DM occurs in $50\lesssim x\lesssim x_f\approx140$. In the right panel of FIG.\ref{IDconstraint2}, we show the  constraints of indirect detection and KE in the $m_X-g_D$ plane. The colored solid lines represent the parameters which can reproduce the relic abundance of DM. The corresponding parameters are chosen to be $m_X/\Lambda'=0.020$ (red),~$0.025$~(magenta),~$0.030$~(green),~$0.035$~(blue),~$0.040$~(purple) by fixing $r_{XZ'}^{1/2}=1.3,~t_\epsilon=10^{-11},~\xi_X=\xi_{Z'}=10^{-5}$. We find that for the mass region, $m_X\geq1$~TeV, $m_X/\Lambda'\gtrsim0.030$ is required for maintaining the KE of DM before freeze-out.

	\section{Conclusion}\label{sect5}
	In this work, we propose a vector dark matter (DM) model in which the DM relic density is determined by the catalyzed freeze-out mechanism. In our model, the DM candidate $X_\mu$ and a catalyst $Z'_\mu\approx C_\mu$ are unified into the dark SU(2)$_D$ gauge fields. The SU(2)$_D$ gauge symmetry is spontaneously broken by VEVs of a doublet and a real triplet scalar fields. Since the catalyst only acquires its mass from the doublet while the DM acquires its mass from both the doublet and triplet, the catalyst is automatically lighter than the DM. The mass condition $3m_{Z'}>2m_x$ for the process $Z'+Z'+Z'\to X^\ast+X$ can also be naturally achieved if the VEVs of the scalar fields satisfy $v_3\lesssim0.56v_2$. Since the catalyzed freeze-out mechanism requires the catalyst to decay after the DM freezes out, we need to introduce a dimension-5 operator $B^{\mu\nu}\Delta^aV^a_{\mu\nu}$ to break a discrete symmetry $G_D$. Such an operator can be easily induced in one loop level by introducing a super heavy fermionic doublet of SU(2)$_D$.
	
	We derive the annihilation cross sections of all the relevant processes, especially, $X^\ast+X\to Z'+Z'$ and $Z'+Z'+Z'\to X^\ast+X$. Then we develop an analytical approximate solution to the Boltzmann equations and compare them to numerical computations. We find that our approximate solution works well so we use them to discuss the constraints from cosmological and astrophysical observations. We provide three benchmark models in which the observed dark matter relic abundance can be achieved. The direct detection constraint is weak in our models since small Higgs-portal couplings can be chosen. However, this model predicts relatively strong DM annihilation cross section, and thus indirect detection experiments can put stringent constraints on it. We find that the $\gamma$-ray spectrum from the Fermi-LAT experiment has excluded the mass region of $m_X<1.2$~TeV for the models with a long-living catalyst. On the other hand, In a model that the catalyst decay during the catalyzed annihilation era, the Fermi-LAT constraint gets looser since a smaller gauge coupling $g_D$  is required by the DM relic abundance. We also find that our model can be tested in the next generation of high energy $\gamma$-ray observations, such as the CTA experiment.
	
	All these discussions are based on an assumption that the DM are kept in kinetic equilibrium (KE) with the thermal bath. However, no concrete mechanism is known to be capable of achieving the KE in such a late time (about $T\sim m_X/10^3$). In the last section, we propose two paradigms which can partially solve the problem. In the first paradigm, DM particle maintains KE with the plasma via $X$-fermions scattering mediated by the Higgs boson. In the second paradigm, we introduce a thermal axion-like particle (ALP), which can collide with DM particles directly, and also plays the role of a mediator of $X$-fermions scattering. In both paradigms, there are parameter spaces allowing DM particles to keep in KE with the plasma until the DM freeze-out. The price we need to pay is that the freeze-out temperature of the catalyst is determined by an extra parameter in both cases. The DM freeze-out via catalyzed annihilation still works, but it implies a freeze-out temperature with $x_f\sim \mathcal{O}(100)$, which is about an order of magnitude higher than that from the original model.

	\acknowledgements
	This work is supported by the National Natural Science Foundation of China (NSFC) under Grant Nos. 11875327 and 11905300, the Fundamental Research Funds for the Central Universities, the Natural Science Foundation of Guangdong Province, and the Sun Yat-Sen University Science Foundation.
	
	%\appendix
	
	\bibliographystyle{JHEP-2-2}
	\bibliography{reference}
\end{document}